\documentclass[10pt,aps,prd,twocolumn,showpacs,superscriptaddress,nofootinbib,nobibnotes,longbibliography,floatfix]{revtex4-1}

\usepackage[utf8]{inputenc}
\usepackage[T1]{fontenc}

\usepackage{bm}
\usepackage{mathtools,amsmath,amssymb,amsfonts,mathrsfs,eucal,graphicx,tensor,csquotes,accents,commath,chngcntr,siunitx}
\usepackage[dvipsnames]{xcolor}
\usepackage[unicode]{hyperref}
\hypersetup{colorlinks=true, citecolor=MidnightBlue,
            linkcolor=MidnightBlue, urlcolor=MidnightBlue, linktocpage=true}
\usepackage[normalem]{ulem}
\usepackage{orcidlink}

\DeclareMathAlphabet{\mathpzc}{OT1}{pzc}{m}{it}
\newcommand{\orcid}[1]{\href{https://orcid.org/#1}{\includegraphics[width=10pt]{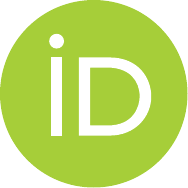}}}

\begin{document}

\title{Quasi-Normal Modes from Bound States: The Numerical Approach}
\author{Sebastian H. V\"olkel \orcid{0000-0002-9432-7690}}
\affiliation{Max Planck Institute for Gravitational Physics (Albert Einstein Institute), D-14476 Potsdam, Germany}
\affiliation{SISSA - Scuola Internazionale Superiore di Studi Avanzati, via Bonomea 265, 34136 Trieste, Italy and INFN Sezione di Trieste}
\affiliation{IFPU - Institute for Fundamental Physics of the Universe, via Beirut 2, 34014 Trieste, Italy}

\email{sebastian.voelkel@aei.mpg.de}

\date{\today}

\begin{abstract}
It is known that the spectrum of quasi-normal modes of potential barriers is related to the spectrum of bound states of the corresponding potential wells. 
This property has been widely used to compute black hole quasi-normal modes, but it is limited to a few ``approximate'' potentials with certain transformation properties for which the spectrum of bound states must be known analytically. 
In this work we circumvent this limitation by proposing an approach that allows one to make use of potentials with similar transformation properties, but where the spectrum of bound states can also be computed numerically. 
Because the numerical calculation of bound states is usually more stable than the direct computation of the corresponding quasi-normal modes, the new approach is also interesting from a technical point of view. 
We apply the method to different potentials, including the P\"oschl-Teller potential for which all steps can be understood analytically, as well as potentials for which we are not aware of analytic results but provide independent numerical results for comparison. 
As a canonical test, all potentials are chosen to match the Regge-Wheeler potential of axial perturbations of the Schwarzschild black hole. 
We find that the new approximate potentials are more suitable to approximate the exact quasi-normal modes than the P\"oschl-Teller potential, particularly for the first overtone. 
We hope this work opens new perspectives to the computation of quasi-normal modes and finds further improvements and generalizations in the future. 
\end{abstract}

\maketitle

\section{Introduction}\label{intro}

Although black hole perturbation theory is in general a non-trivial field of research, it has provided some surprisingly simple results. 
In general relativity, as well as in many modified theories of gravity, it is possible to derive so-called master equations that break down the full scale of the problem into finding complex frequency eigenvalues of effective potentials in a one-dimensional Schr\"odinger equation or modifications of it \cite{Regge:1957td,Vishveshwara:1970cc,Zerilli:1970se,Teukolsky:1973ha}. 
Non-rotating black holes are significantly easier to treat and it is quite generic to find single or coupled wave equations also in modified gravity. 
However, similar results for rotating black holes are limited to general relativity and very few specific theories for which the complicated calculations could be carried out or are limited to small spins, see Refs.~\cite{Cano:2020cao,Cano:2021myl,Wagle:2021tam,Pierini:2021jxd,Pierini:2022eim} for recent developments. 

The close relation to quantum mechanics immediately calls for similar methods to solve the final eigenvalue problem and the literature on adopted as well as new methods has grown immensely, see e.g. Refs.~\cite{Kokkotas:1999bd,Nollert:1999ji,Berti:2009kk,Konoplya:2011qq,Pani:2013pma} for reviews. 
Methods to compute black hole quasi-normal modes range from purely analytic, semi-analytic and fully numerical ones. 
The question of which method to choose from ultimately depends on the specifics of the problem and the desired insights that should be gained. 
Methods that provide quasi-normal mode frequencies with pristine precision, such as the Leaver method \cite{Leaver:1985ax}, do not provide analytic results and might be difficult to adjust for new potentials. 
Analytic approaches, such as the application of the Wentzel-Kramers-Brillouin (WKB) method, can provide analytic results, but those are approximate and may not apply to all parts of the quasi-normal mode spectrum \cite{Schutz:1985km,Iyer:1986np,Iyer:1986nq,Kokkotas:1988fm,Seidel:1989bp,Guinn:1989bn,Kokkotas:1993ef,Andersson_1993,Matyjasek:2017psv}. 
For example, the higher order WKB method can be very precise for the first few overtones, but it qualitatively fails for large overtones. 
See Ref.~\cite{bender1999advanced} for a standard textbook on the topic or Ref.~\cite{Dunham:1932zz} for an early application to related problems in quantum mechanics. 
A recent review that is more specific to the application of the WKB method to black holes can be found in Ref.~\cite{Konoplya:2019hlu}. 

One method that is particularly insightful with respect to standard quantum mechanics is the inverted potential method proposed by Mashhoon \cite{Mashhoon:1982im}, which is based on earlier work of Heisenberg \cite{Heisenberg:1946ytd} and was further explored by him and collaborators in Refs.~\cite{BLOME1984231,Ferrari:1984ozr,Ferrari:1984zz}. 
It establishes a mapping between the bound states of a potential well and the quasi-normal modes of the corresponding potential barrier if certain transformation properties exist and the bound states are known analytically, by e.g. the factorization method Ref.~\cite{1951RvMP...23...21I}. 
While the complicated potentials of black holes do not allow for the analytic computation of bound states, the method has been applied to approximate potentials for which analytic results are known, e.g. the Eckart potential \cite{Eckart:1930zza}, the P\"oschl-Teller (PT) potential \cite{Poschl:1933zz}, and more recently a modified PT potential \cite{Churilova:2021nnc}. 
The simplicity and ease of use of this method made it very popular in the literature, but as with other analytic methods, it has certain limitations. 
The main limitation is that the method cannot be readily applied to potentials for which the spectrum of bound states can only be computed numerically. 
Therefore the number of suitable potentials in the standard approach is limited. 

Since then extensions of the original idea have been studied. 
In Ref.~\cite{Zaslavsky:1991ug} it has been suggested to use an anharmonic oscillator potential to compute quasi-normal modes from bound states in a perturbative way and in Ref.~\cite{Galtsov:1991nwq} a perturbative, complex valued WKB matrix approach has been presented. 
Based on the Bender-Wu method \cite{Bender:1973rz} and subsequent work \cite{Sulejmanpasic:2016fwr}, it was shown in Ref.~\cite{Hatsuda:2019eoj} that a similar perturbative treatment, based on using an anharmonic oscillator potential, Pad\'e approximants and the Borel summation, allows to compute quasi-normal modes from bound states with very high precision. 
This idea was further improved in Ref.~\cite{Matyjasek:2019eeu}, which demonstrates that it also allows for more precise calculation of overtones. 

The purpose of this work is to extend the method to more general potentials for which bound states are obtained numerically, which does not require one to limit oneself to potentials that can be locally represented by a Taylor expansion around their minimum. 
The key idea behind the numerical extension is to compute an analytic representation of the spectrum of bound states by using the numerical results and then apply the necessary transformations to it. 
We use a Taylor expansion of the bound state spectrum around a given parameter choice, but other representations can in principle be used as well. 
Limiting numerical calculations to the bound state problem is rewarding because it is generically more stable and in practice much easier to perform than those of quasi-normal modes. 
We demonstrate the performance of the new method by applying it first to the PT potential, for which all steps can be verified analytically. 
Then we study two potentials for which we are not aware of analytic results in the literature. 
In all cases we choose the parameters of the potential such that they correspond to an approximation of the Regge-Wheeler (RW) potential that describes the axial perturbations of the Schwarzschild black hole \cite{Regge:1957td}, which we regard as the default benchmark test of the method. 
Our results demonstrate that the method is simple to use and provides more precise results than using the PT potential as known analytic approximation. 

The paper is structured as follows. 
In Sec.~\ref{method} we first review the analytic approach and then outline the new method. 
The application of the method to different potentials is demonstrated in Sec.~\ref{applications}. 
We discuss our findings and provide remarks for further extensions of the method in Sec.~\ref{discussion}. 
Finally, our conclusions can be found in Sec.~\ref{conclusions}. 
In Appendix~\ref{appendix} we provide additional material for the shooting method. 
Throughout this work we use units in which $G = c = 1$. 

\section{Method}\label{method}

\subsection{Bound States and Quasi-Normal Modes}

The starting point of this work is to consider the standard Schr\"odinger equation
\begin{align}\label{schrodinger}
\frac{\text{d}^2}{\text{d}x^2} \Psi(x) + \left[E_n-V(x, P) \right] \Psi(x) = 0,
\end{align}
where $V(x,P)$ is a potential with some parameter(s) $P$ and $E_n$ is the corresponding spectrum of eigenvalues for a given choice of boundary conditions. 
Very qualitatively, if $V(x, P)$ describes a potential well, the physical boundary conditions for bound states are those for which $\Psi(x) \rightarrow 0$ for $x \rightarrow \pm \infty$. 
Depending on the properties of $V(x,P)$, this can give rise to a finite or infinite set of eigenvalues. 
If however $V(x, P)$ describes a potential barrier, the suitable boundary conditions depend on the application in mind. 

Quasi-normal modes in the context of black holes are usually defined as purely outgoing solutions of the time dependent wave equation, which in the time independent Eq.~\eqref{schrodinger} correspond to diverging solutions for $\Psi(x)$ for $x \rightarrow \pm \infty$. 
For an extended introduction to quasi-normal modes and other techniques to compute them we refer the interested reader to Refs.~\cite{Kokkotas:1999bd,Nollert:1999ji,Berti:2009kk,Konoplya:2011qq,Pani:2013pma} and continue with reviewing a few basics in the following. 

The two potentials that describe gravitational perturbations around the Schwarzschild black hole in general relativity are known as the Regge-Wheeler potential \cite{Regge:1957td} and the Zerilli potential \cite{Zerilli:1970se}. 
It can be shown that both potentials are isospectral to each other, although their analytic structure is different \cite{Chandrasekhar:1985kt,Glampedakis:2017rar}. 
In the following we continue with the RW potential, which is given by
\begin{align}\label{V_RWo}
V_\text{RW} (r(x), M, l) = \left(1 - \frac{2M}{r} \right) \left(\frac{l(l+1)}{r^2} - \frac{6 M}{r^3} \right).
\end{align}
Here $M$ is the mass of the black hole and $l(l+1)$ is a separation constant with $l \geq 2$. 
The familiar form of the Schr\"odinger equation with potential term only appears in the so-called tortoise coordinate $x$, defined via
\begin{align}\label{tortoise}
x = r + 2M\ln \left(\frac{r}{2M} - 1\right),
\end{align}
which makes a full analytic treatment more involved. 
Note that exact analytic solutions have been found in terms of confluent Heun functions \cite{Fiziev:2005ki}, but the quasi-normal mode spectrum cannot be written in terms of simple functions. 

\subsection{Review of the Analytic Method}

In Refs.~\cite{Mashhoon:1982im,BLOME1984231,Ferrari:1984ozr,Ferrari:1984zz} the following structure of the Schr\"odinger Eq.~\eqref{schrodinger} was noticed and utilized. 
If one considers the transformation $x\rightarrow - i x$ and is able to transform the original parameters $P$ of the potential to a new set of parameters $P^\prime = \pi(P)$ such that
\begin{align}\label{potential_condition}
V(x, P) = V(-ix, P^\prime), 
\end{align}
one can use the spectrum of bound states $\Omega^2(P) \equiv -E_n(P)$ of the potential well to compute the quasi-normal modes of the potential barrier via 
\begin{align}\label{bound_to_qnm}
\omega_n(P) \equiv \Omega_n(\pi^{-1}(P)).
\end{align}
The success of the method depends on whether the analytic form of the bound states $E_n(P)$ is known. 
Because an application to complicated potentials usually does not allow for an analytic computation of the spectrum, but the analytic form is needed to apply Eq.~\eqref{bound_to_qnm}, the method cannot be used. 
In the following we show how this major shortcoming can be circumvented and how the method can be used for potentials where the bound states can be computed numerically. 

\subsection{Numerical Bound State Method}

The recipe of the numerical bound state method is as follows. 
It is assumed that one has a precise numerical method available to compute the bound states $E_n(P)$ as function of a given set of parameters $P$. 
One straightforward approach is the shooting method, see e.g. Ref.~\cite{Press2007} for an overview. 
The method is based on integrating $\Psi(x)$ as initial value problem from two distant points for a given choice of $E$ and computing the Wronskian of both solutions at an intermediate point. 
This process can be formulated as root finding problem, because the Wronskian vanishes when the initial guess for $E$ is an eigenvalue $E_n$. 
Because the method is widely used we refer to Appendix \ref{appendix_bound} for more details. 
In Appendix \ref{appendix_qnm} we review the direct shooting method for the quasi-normal mode case and provide supplementary information. 

Next it is assumed that $E_n$, for each $n$ respectively\footnote{We suppress the $n$ dependency on each $T_k$ for simplicity.}, can be represented by a Taylor series around a given parameter set $P^{0}$ 
\begin{align}
E_n(P) \approx \sum_{k}{} T_k.
\end{align}
Up to second order the series can be written in the compact form
\begin{align}\label{taylor_En}
E_n(P) \approx T_0 + T_1 + T_2, 
\end{align}
with 
\begin{align}
T_0 &= E^{0}_n, \\
T_1 &= \text{grad}(E_n)\big|_{P=P^{0}} \left(P - P^{0}\right),\\
T_2 &= \frac{1}{2}  \left(P - P^{0}\right) \text{H}(E_n)\big|_{P=P^{0}} \left(P - P^{0}\right).
\end{align}
Here the gradient and Hessian operators are defined as
\begin{align}
\left[ \text{grad} (E_n) \right]_i   &= \frac{\text{d} E_n}{\text{d} P_i},
\\
\left[ \text{H}(E_n)\right]_{ij} &= \frac{\text{d}^2 E_n}{\text{d} P_i \text{d} P_j}.
\end{align}
All derivatives can be obtained numerically, e.g. in terms of higher order finite differences, by using the numerical method to compute bound states in the vicinity of $P^{0}$. 
Note that it is in principle also possible to extend the multi-dimensional Taylor series to higher order. 

As last step we consider the transformations. 
If the inverse transformation $\pi^{-1}(P)$ is known, one can simply compute the approximate form of the quasi-normal modes at any $P^{0}$ from inserting it in the Taylor ansatz Eq.~\eqref{taylor_En}. 
Note that the transformations typically extend the parameters to the complex plane and that they do not have to be close to the expansion point of the Taylor series. 
In this case the convergence of the Taylor series needs to be carefully studied and higher order terms may become necessary. 
We emphasize that the validity of the Taylor series, or any other expression found for the bound state spectrum, is crucial and non-trivial for arbitrary potentials. 

An important simplification of the multi-dimensional Taylor series can be obtained if the inverse transformation $\pi^{-1}$ is just the identity. 
In that case all terms associated with $(P_i - P^{0}_i)$ are zero when evaluated at the expansion point of the Taylor series, while the non-trivial transformations give non-zero contributions
\begin{align}
 \left( \pi^{-1}(P_i) - P^{0}_i\right)\big|_{P_i = P^{0}_i}
  \Big\{\begin{array}{lr}
        =0, & \text{for } \pi^{-1}(P_i) = P_i, \\
        \neq 0, & \text{for }  \pi^{-1}(P_i) \neq P_i.
        \end{array}
\end{align}
This has the important and practical simplification that the higher order terms of the multi-dimensional Taylor series that one actually has to consider only depend on the variables whose transformations are non-trivial. 
Those parameters may change the spectrum at the zeroth order term, but because the quasi-normal modes are computed exactly for $\pi^{-1}(P^0)$ all higher order terms vanish. 

\section{Applications}\label{applications}

In this section we apply the numerical bound state method to different potentials. 
In Sec.~\ref{V_RW} we first discuss the transformation properties of the RW potential and how approximate potentials are used. 
In Sec.~\ref{V_PT} we then apply the numerical method to the well known PT potential, because the results are known analytically and all steps can therefore be understood carefully. 
In Sec.~\ref{V_BW} and Sec.~\ref{V_MIX} the method is applied to the Breit-Wigner potential and a piecewise combination. 

All shown derivatives in this section are computed using finite differences with $9$ or $11$ point stencils \cite{fdcc} and different step sizes indicated in the caption of each table and figure. 
The step-size $h$ is defined by a dimensionless factor $\epsilon$
\begin{align}\label{step_size}
h = \epsilon P,
\end{align}
where $P$ is the only non-trivial parameter of each potential. 
Its numerical value is chosen to fit the $M=1$ and $l=2$ RW potential as described in Sec.~\ref{V_RW}. 
For comparison we also provide the quasi-normal modes for the BW and mixed potential using the direct shooting method outlined in Appendix \ref{appendix_qnm}. 

\subsection{Regge-Wheeler Potential}\label{V_RW}

In contrast to the approximate potentials studied in the literature, finding the transformations $\pi(P)$ for the RW potential is less obvious, but they were already reported in Refs.~\cite{Ferrari:1984ozr,Ferrari:1984zz}. 
By introducing a new parameter $\lambda$ as overall factor in the original potential Eq.~\eqref{V_RWo}
\begin{align}
V^{({\lambda)}}_\text{RW} (r(x), M, l, \lambda) = \lambda V_\text{RW} (r(x), M, l),
\end{align}
one can fulfill the necessary condition Eq.~\eqref{potential_condition} for $\lambda \rightarrow -\lambda$. 
The original potential is included as trivial case for $\lambda=1$ and the full set of transformations is thus given by 
\begin{align}
\pi(M) &= -iM, \quad \pi(l) = l, \quad \pi(\lambda) = -\lambda, \\
\pi^{-1}(M) &= iM, \quad \pi^{-1}(l) = l, \quad \pi^{-1}(\lambda) = -\lambda.
\end{align}
However, because the spectrum of bound states of the inverted RW potential is not known analytically, the original works are limited to the application of the harmonic oscillator, the PT potential or the Eckart potential as approximations \cite{Mashhoon:1982im,BLOME1984231,Ferrari:1984ozr,Ferrari:1984zz}. 
For these potentials one simply demands that the value of the maximum, as well as the value of the second derivative at the maximum, agree with those of the RW potential. 
One therefore demands that the approximate potential must locally match the RW potential around its maximum. 
Because all approximate potentials are matched at the peak of the potential, and it is known that the fundamental quasi-normal is tightly related to it, the approximate potentials are useful to estimate the fundamental mode $n=0$, but may fail for overtones $n>0$. 
In the following we adopt for simplicity the notation that the tortoise coordinate is labeled with $x$ but shifted by whatever the exact location of the RW potential is, such that the maximum is always at $x=0$. 
This translation does not change the quasi-normal modes spectrum.

\begin{figure}
\includegraphics[width=1.0\linewidth]{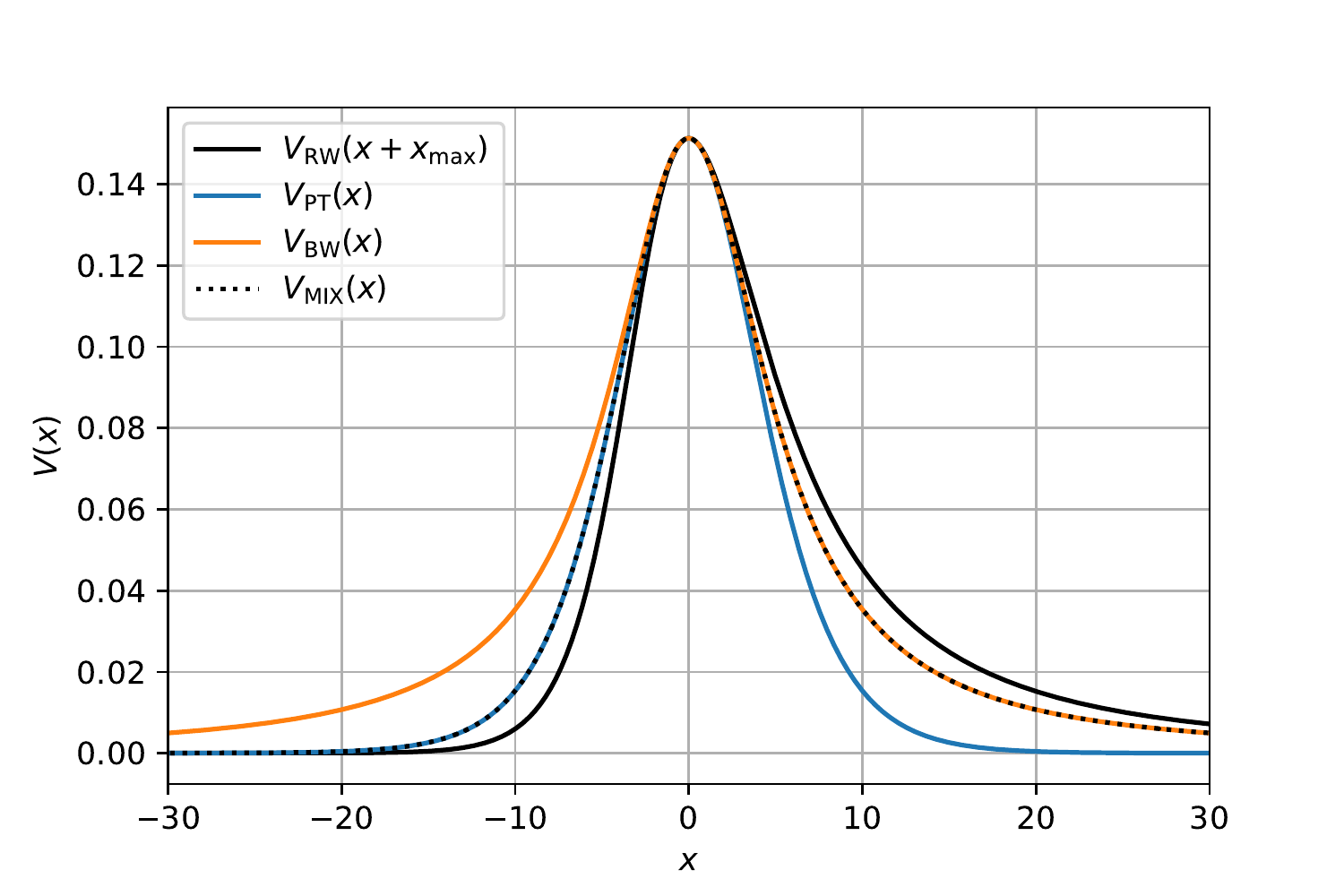}
\\
\includegraphics[width=1.0\linewidth]{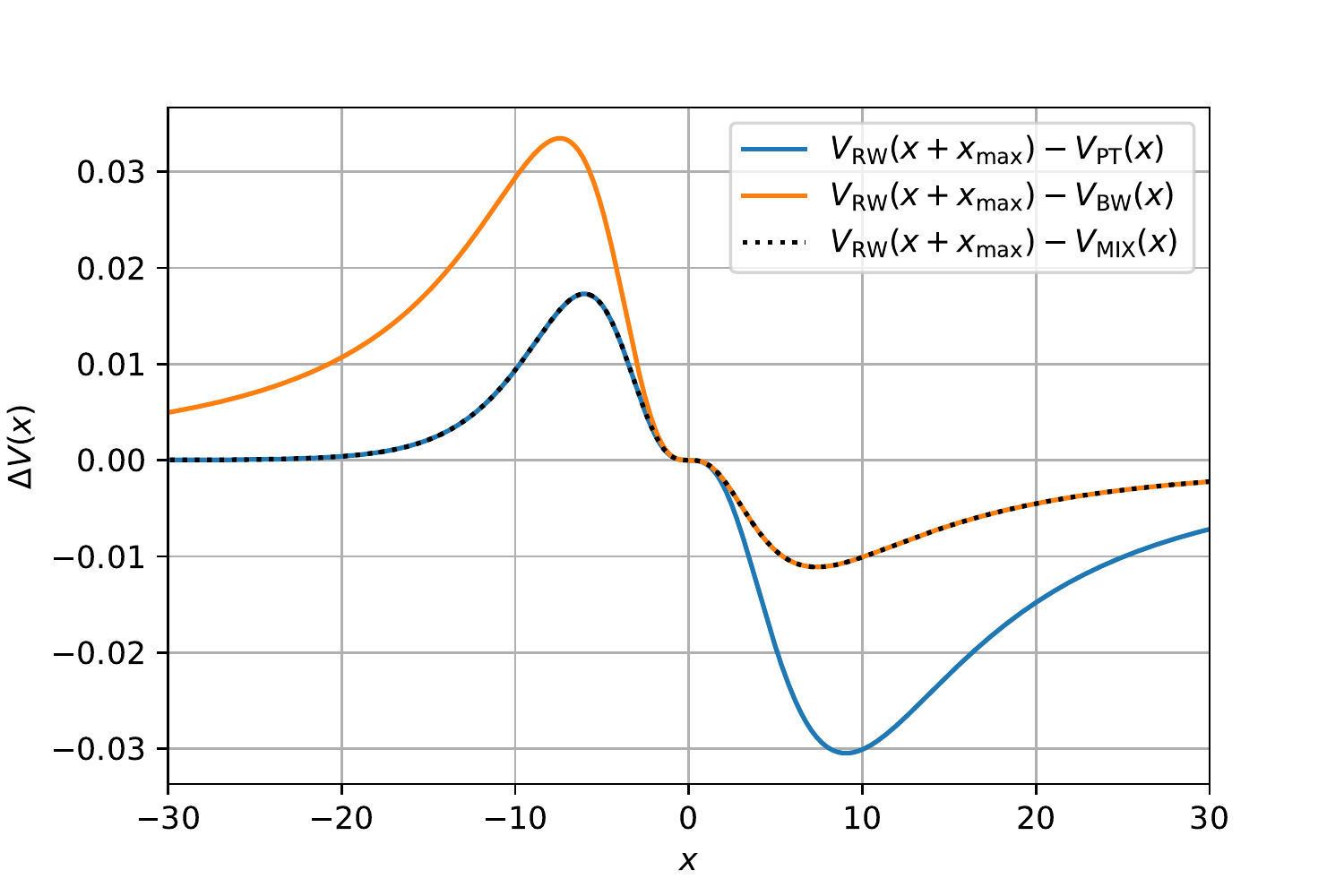}
\caption{In this figure we compare the RW potential for $M =1$ and $l=2$ (black) with the PT potential (blue), the BW potential (orange) and the mixed potential (black dotted). The parameters of the approximate potentials have been chosen to agree with the RW at its peak. In the top panel we show the different potentials, while the bottom panel shows the difference of a given potential with respect to the RW potential.}
\label{V_plots}
\end{figure}

Before introducing each potential in the following sections, we show all of them for parameters chosen to match the RW potential for $M=1$ and $l=2$ in Fig.~\ref{V_plots}. 
It is evident that neither the PT potential Eq.~\eqref{PT_potential} nor the BW potential Eq.~\eqref{BW_potential} correctly capture both asymptotic behaviours of the RW potential simultaneously, but only one of them. 
As a ``mixed'' potential Eq.~\eqref{MIX_potential} we introduce a piecewise combination, which shows better agreement. 

\subsection{P\"oschl-Teller Potential}\label{V_PT}

The original method has been applied to black holes in Refs.~\cite{Mashhoon:1982im,Ferrari:1984ozr,Ferrari:1984zz} by using the PT potential as an approximation for the more complicated black hole potentials. 
The PT potential is defined in slightly different ways in the literature, but in the following we will use the form
\begin{align}\label{PT_potential}
V(x, P) = \frac{V_0}{\cosh^2(\alpha x)}.
\end{align}
Here $P=(V_0, \alpha)$ are the parameters describing the ``height/depth'' and ``curvature'' at the maximum/minimum. 
The transformations $\pi_i(P_i)$ and their inverses are simply given by
\begin{align}
\pi_{V_0}(V_0) &= V_0, \quad \pi_{\alpha}(\alpha) = i \alpha,
\\
\pi^{-1}_{V_0}(V_0) &= V_0,\quad  \pi^{-1}_{\alpha}(\alpha) = -i \alpha.
\end{align}
The spectrum of bound states is given by $E_n = -\Omega_n^2$ with
\begin{align}\label{Omega_PT}
\Omega_n(V_0, \alpha) = \alpha \left[-\left(n+\frac{1}{2}\right) + \left(\frac{1}{4} + \frac{V_0}{\alpha^2} \right)^{1/2} \right].
\end{align}
Using the inverse transformations, the spectrum of quasi-normal modes is given by
\begin{align}\label{omega_PT}
\omega_n(V_0, \alpha) = \pm \left(V_0 -\frac{\alpha^2}{4} \right)^{1/2} + i \alpha \left(n+\frac{1}{2}\right).
\end{align}
Due to a different choice for the sign convention for the Fourier transform used in Refs.~\cite{Mashhoon:1982im,Ferrari:1984ozr,Ferrari:1984zz} with respect to the rest of this work, we continue using Eq.~\eqref{omega_PT} with a negative imaginary part. 
Because the transformation of $V_0$ is just the identity, the relevant part of the Taylor series of the spectrum needed to apply the inverse transformation of bound states only depends on $\alpha$ and can thus in principle be easily computed even for higher orders than $2$. 
The practical limitation to include very high derivatives is the numerical precision required to compute them numerically from the solutions of the bound state boundary value problem. 

In the following we compare the results obtained with the numerical bound state method with the exact analytic results. 
Because the bound state spectrum is known analytically, we can compare the numerical results for the Taylor expansion of Eq.~\eqref{Omega_PT} around $P$ evaluated at $\pi^{-1}(P)$ with the analytic ones for a given order of the Taylor expansion. 
To demonstrate the convergence of the quasi-normal modes as function of the Taylor order $k$, we define 
\begin{align}\label{rel_err_PT_re}
\delta_k(\omega_\mathrm{re}) &\equiv \frac{\omega_{k, \textrm{re}}-\omega_\mathrm{re}^\mathrm{Exact}}{\omega_\mathrm{re}^\mathrm{Exact}},\\
\delta_k(\omega_\mathrm{im}) &\equiv \frac{\omega_{k, \textrm{im}}-\omega_\mathrm{im}^\mathrm{Exact}}{\omega_\mathrm{im}^\mathrm{Exact}}.\label{rel_err_PT_im}
\end{align}
Here the index ``re/im'' indicates the real or imaginary part and the label ``Exact'' indicates the analytic value.

\begin{figure}
\includegraphics[width=1.0\linewidth]{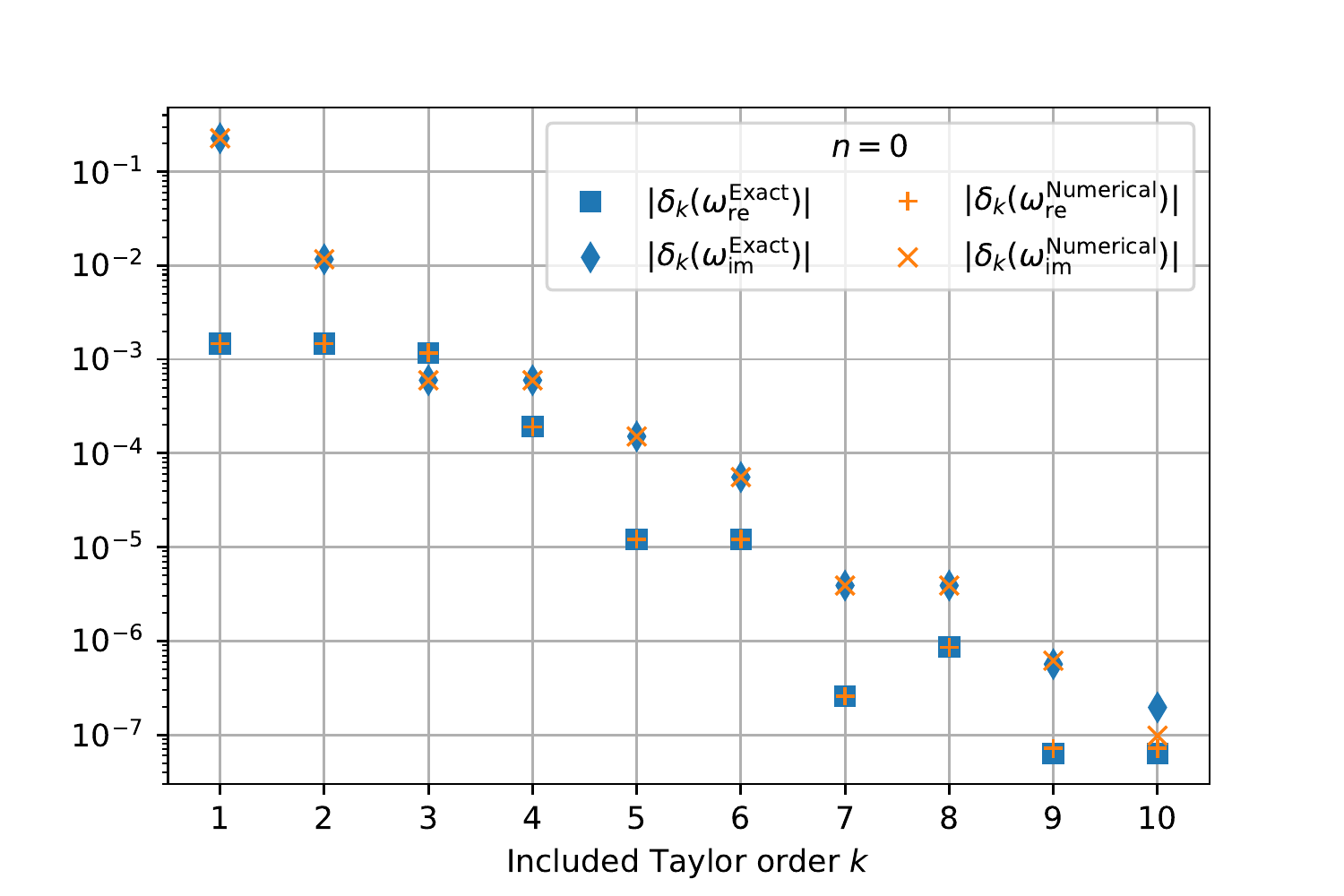}
\includegraphics[width=1.0\linewidth]{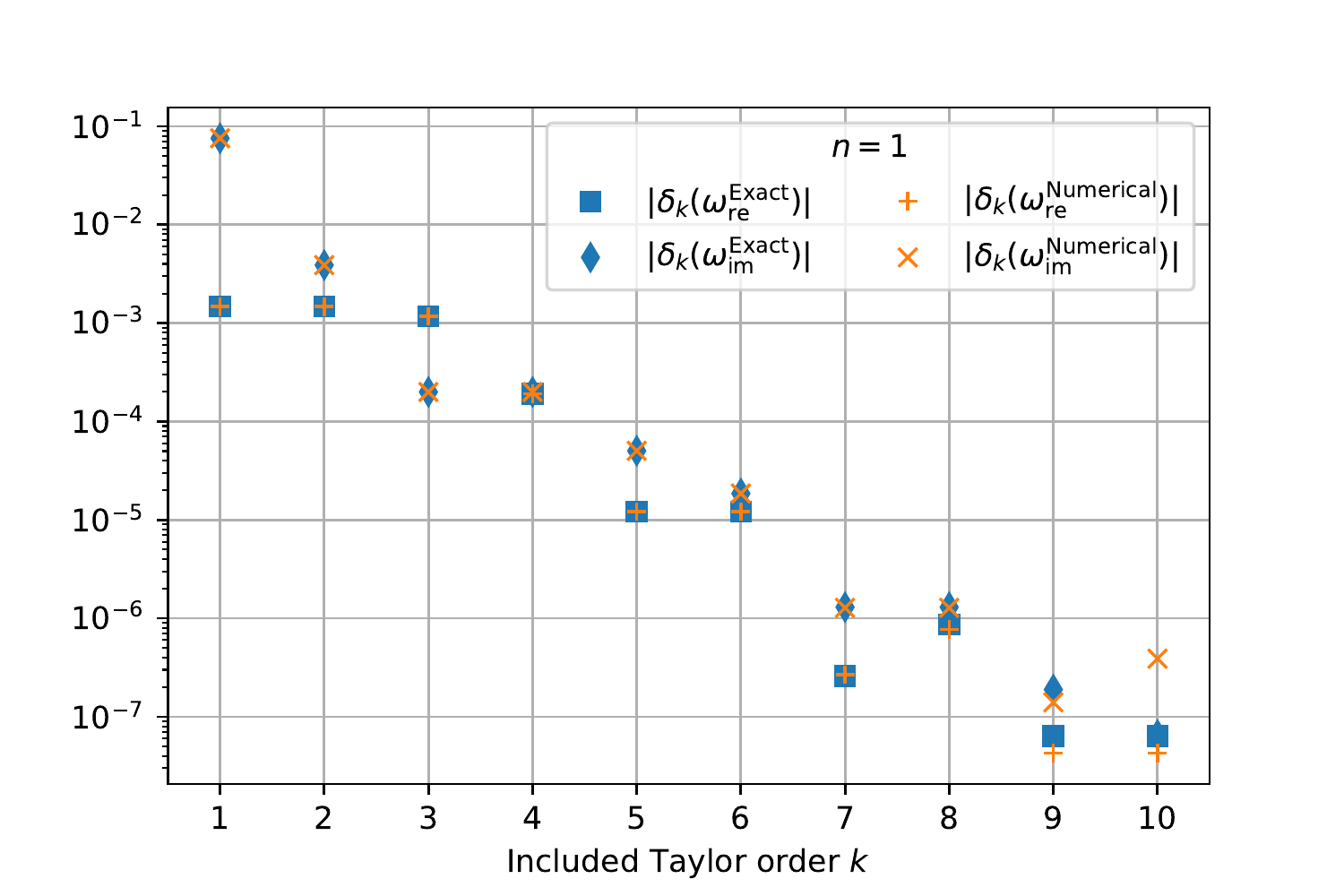}
\caption{Here we show the error defined in Eq.~\eqref{rel_err_PT_re} and Eq.~\eqref{rel_err_PT_im} for the exact (blue) and numerical (orange) results for including different orders $k$ in the Taylor series for the PT potential spectrum for $n=0$ (top panel) and for $n=1$ (bottom panel). 
The real and imaginary parts are shown with different symbols. 
The numerical values for the individual quasi-normal modes can be found in Table \ref{w_PT_table}. 
}
\label{fig_PT_Taylor}
\end{figure}

\begin{figure}
\centering
\includegraphics[width=1.0\linewidth]{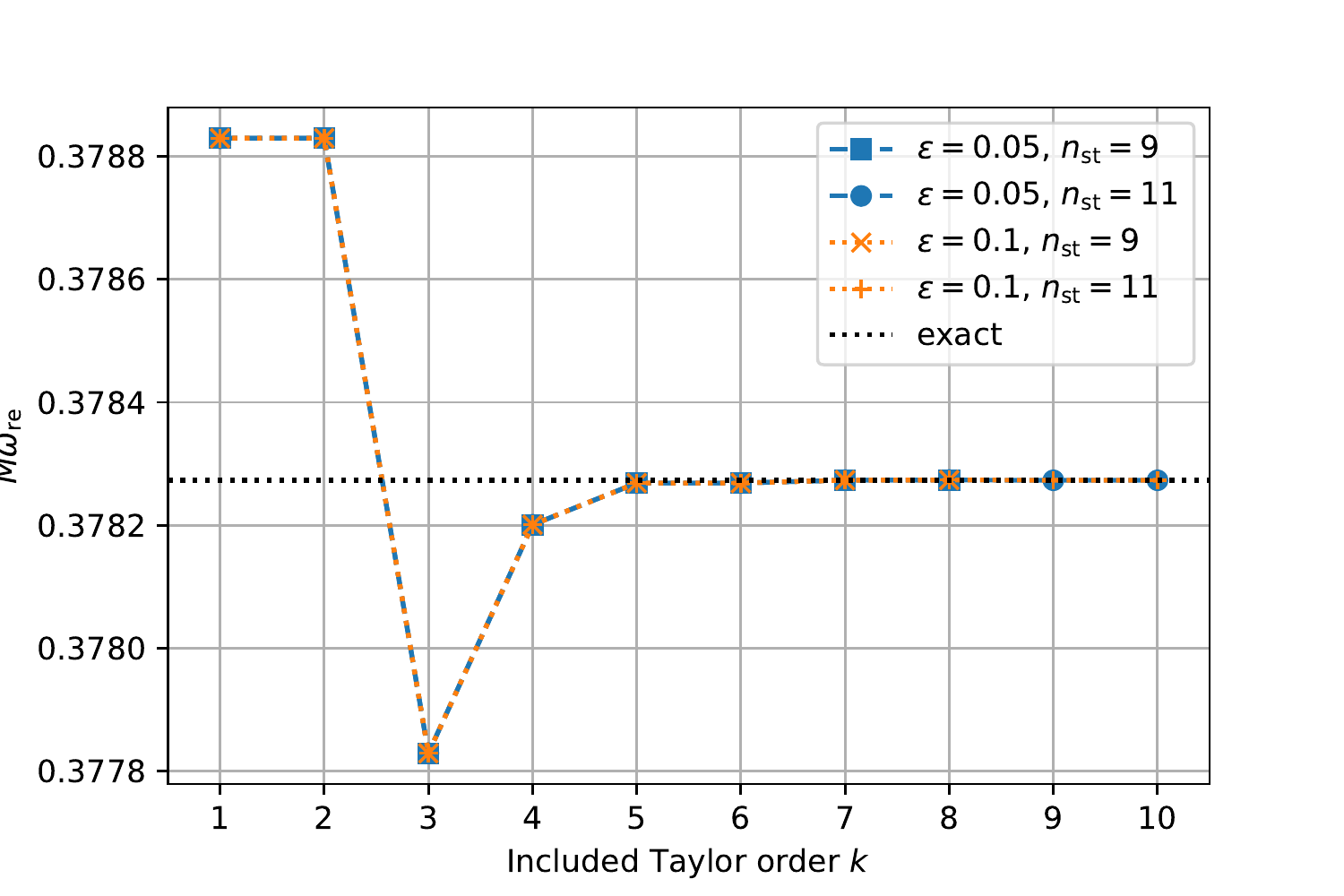}  
\includegraphics[width=1.0\linewidth]{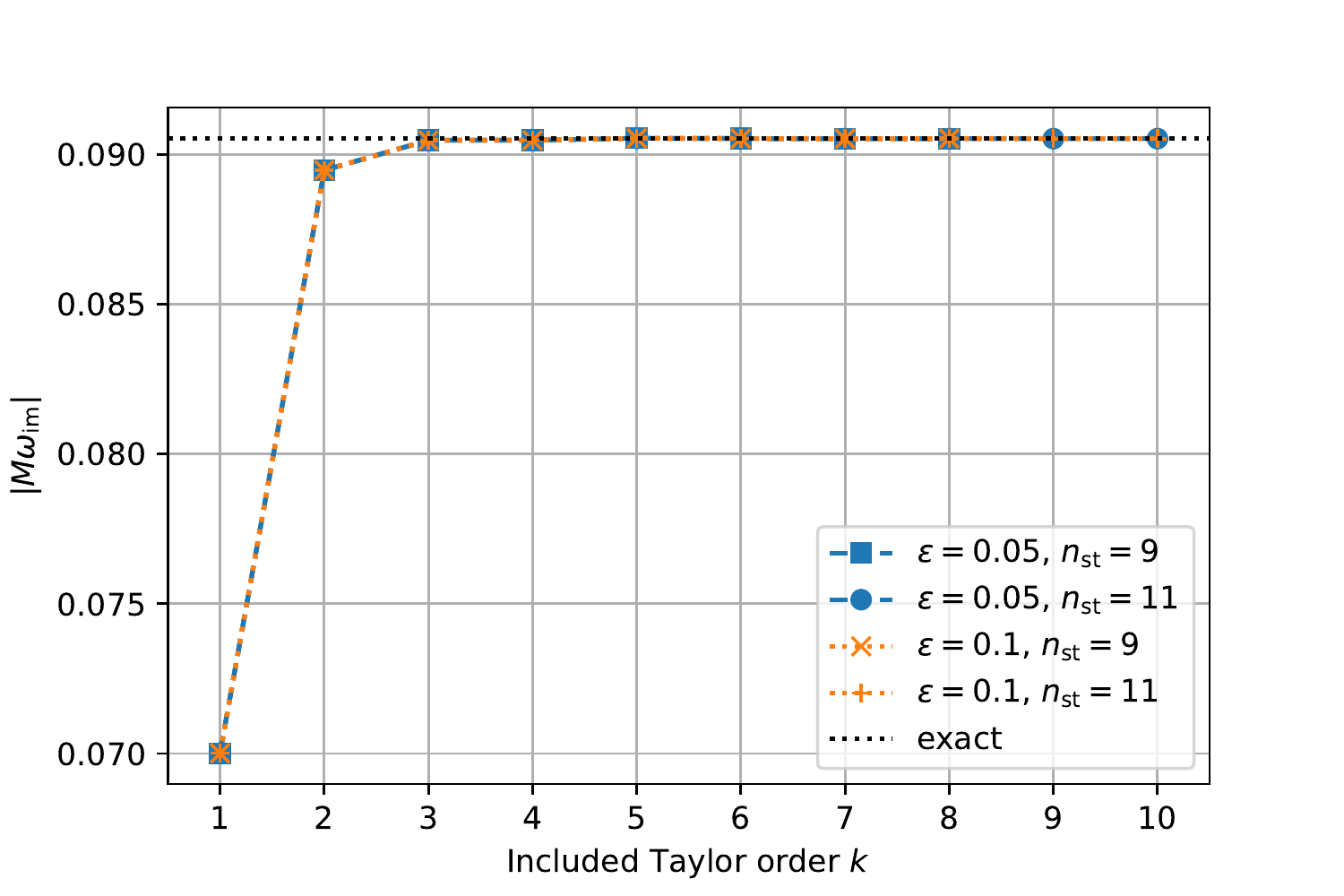}
\caption{Here we show the real part (left panel) and imaginary part (right panel) for the $n=0$ quasi-normal mode computed for the PT potential for different step-sizes $\epsilon = (0.05, 0.1)$ and stencils of $n_\mathrm{st}=(9,11)$. The analytic values are indicated as black dotted line.}
\label{fig_PT_w_0}
\end{figure}

\begin{figure}
\centering
\includegraphics[width=1.0\linewidth]{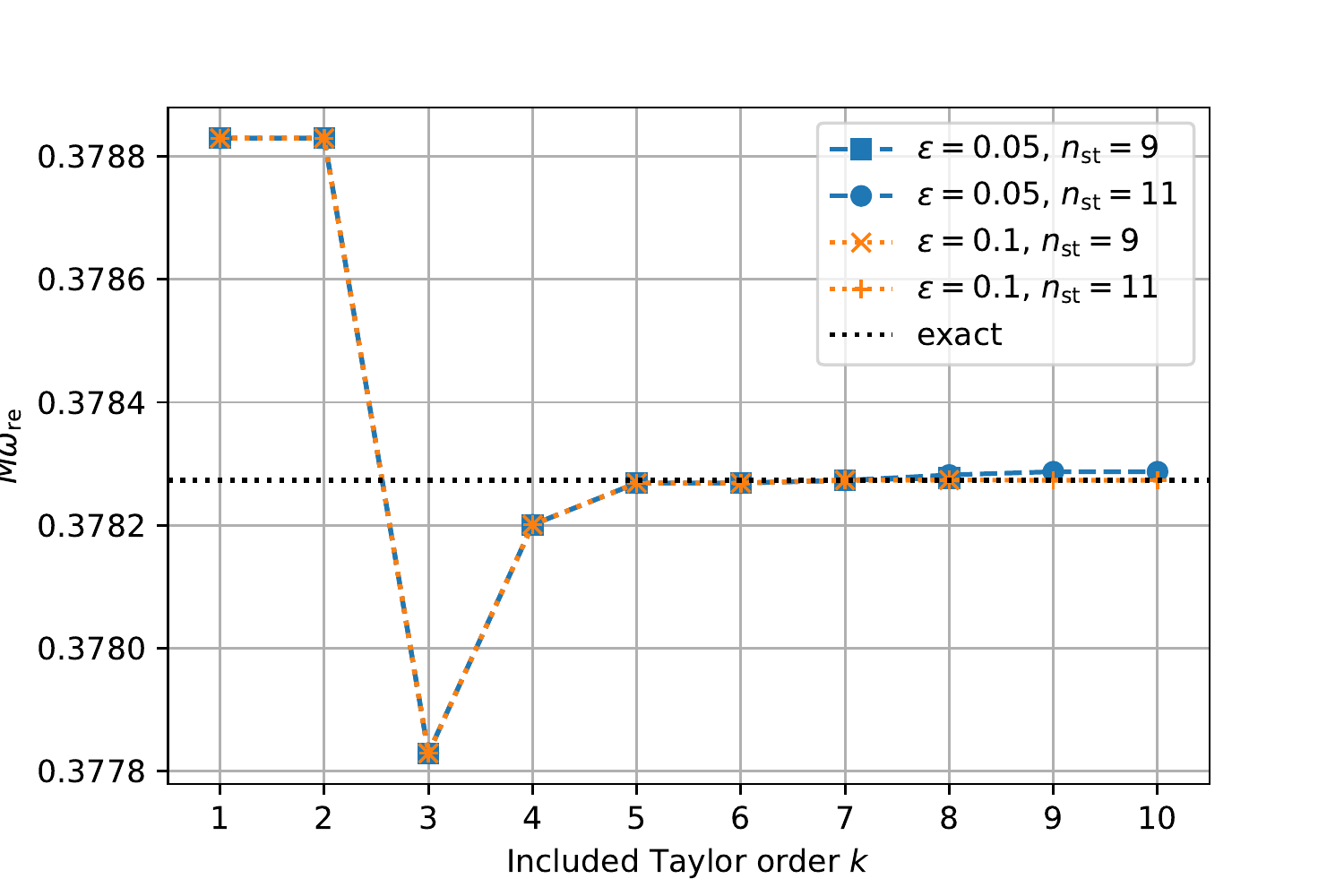}  
\includegraphics[width=1.0\linewidth]{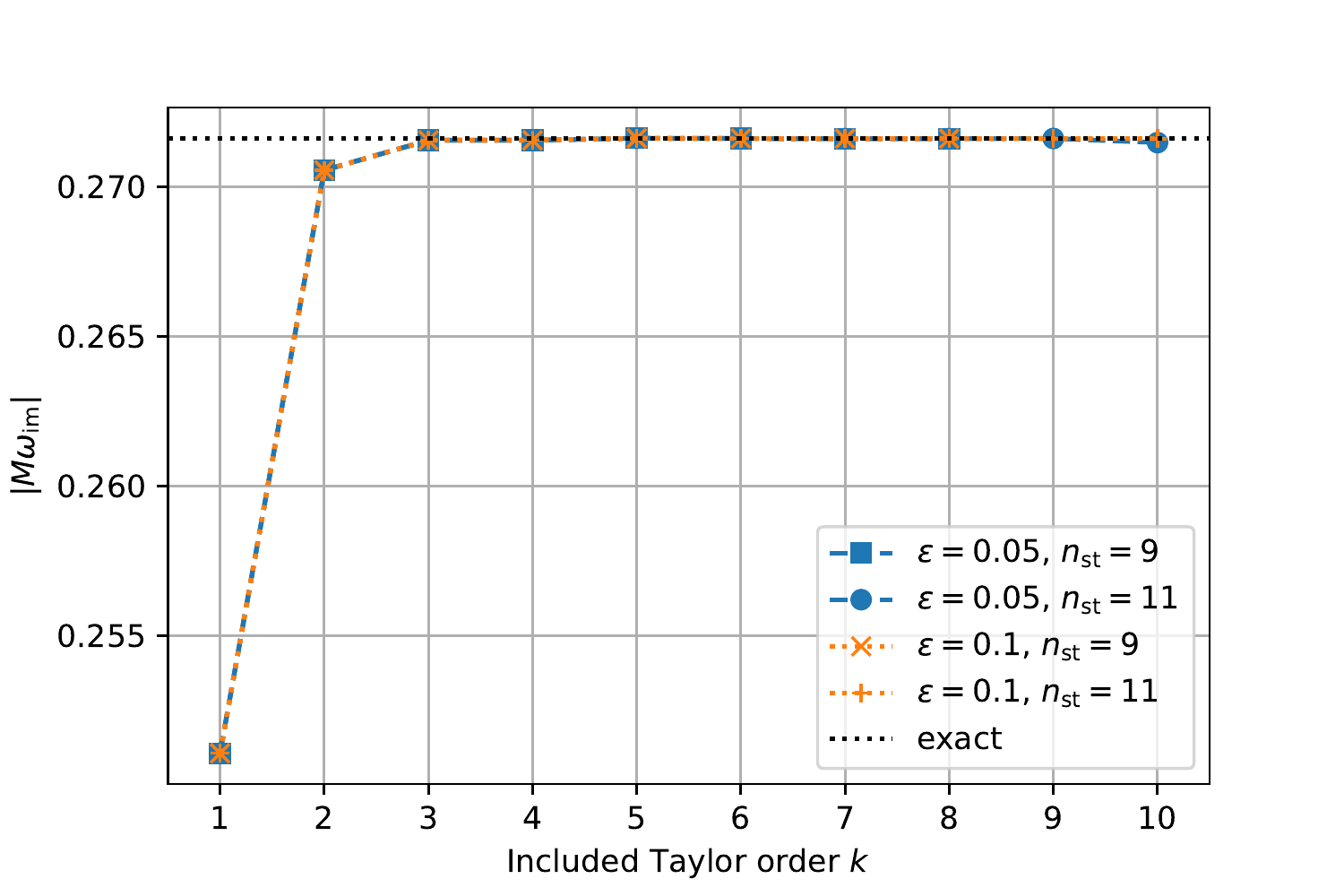}  
\caption{Here we show the real part (left panel) and imaginary part (right panel) for the $n=1$ quasi-normal mode computed for the PT potential for different step-sizes $\epsilon = (0.05, 0.1)$ and stencils of $n_\mathrm{st}=(9,11)$. The analytic values are indicated as black dotted line.}
\label{fig_PT_w_1}
\end{figure}

In Fig.~\ref{fig_PT_Taylor} we show this error function when applied to the exact Taylor series expansion, as well as the numerically computed one. 
It can be seen that the agreement between the analytic and the numerical method is excellent and starts to slightly deviate at the $k=9$ Taylor order. 
At the same time Fig.~\ref{fig_PT_Taylor} also demonstrates that using the Taylor series, be it analytically or numerically computed, yields very accurate approximations for the true quasi-normal modes. 
It is interesting to note that the real part is already captured at $0.1\,\%$ level for $k=1$ and only starts to improve significantly around $k>=4$. 
In contrast, the imaginary is badly approximated at $k=1$ but already at $1\,\%$ level for $k=2$. 
Overall these results show that the $n=0$ and $n=1$ quasi-normal modes can be computed with the numerical bound state method at very high accuracy and that the exact Taylor series has good convergence properties. 
All values used to compute the relative errors are also listed in Table \ref{w_PT_table}. 
In Fig.~\ref{fig_PT_w_0} and Fig.~\ref{fig_PT_w_1} we show the quasi-normal mode frequencies computed for either using 9 or 11 point stencils for the finite difference scheme, as well as for different step-sizes. 
It is evident that the agreement is very good for $n=0$ for all shown Taylor orders, while there are small deviations for very high orders for the real part for $n=1$. 

\begin{table}
\[
\begin{array}{cccc}
\hline
 n & k & M \omega^\mathrm{Exact}_{k} & M \omega^\mathrm{Numerical}_{k} \\
\hline
 0 & 1  & 0.37883 -i 0.07001 & 0.37883 -i 0.07001 \\
   & 2  & 0.37883 -i 0.08948 & 0.37883 -i 0.08948 \\
   & 3  & 0.37783 -i 0.09048 & 0.37783 -i 0.09048 \\
   & 4  & 0.37820 -i 0.09048 & 0.37820 -i 0.09048 \\
   & 5  & 0.37827 -i 0.09055 & 0.37827 -i 0.09055 \\
   & 6  & 0.37827 -i 0.09054 & 0.37827 -i 0.09054 \\
   & 7  & 0.37827 -i 0.09053 & 0.37827 -i 0.09053 \\
   & 8  & 0.37827 -i 0.09053 & 0.37827 -i 0.09053 \\
   & 9  & 0.37827 -i 0.09053 & 0.37827 -i 0.09053 \\
   & 10 & 0.37827 -i 0.09053 & 0.37827 -i 0.09053 \\
   & \mathrm{`` \infty "}& 0.37827 - i 0.09053 & \\
\hline
 1 & 1  & 0.37883 -i 0.25107 & 0.37883 -i 0.25107 \\
   & 2  & 0.37883 -i 0.27054 & 0.37883 -i 0.27054 \\
   & 3  & 0.37783 -i 0.27154 & 0.37783 -i 0.27154 \\
   & 4  & 0.37820 -i 0.27154 & 0.37820 -i 0.27154 \\
   & 5  & 0.37827 -i 0.27161 & 0.37827 -i 0.27161 \\
   & 6  & 0.37827 -i 0.27160 & 0.37827 -i 0.27160 \\
   & 7  & 0.37827 -i 0.27160 & 0.37827 -i 0.27160 \\
   & 8  & 0.37827 -i 0.27160 & 0.37827 -i 0.27160 \\
   & 9  & 0.37827 -i 0.27160 & 0.37827 -i 0.27160 \\
   & 10 & 0.37827 -i 0.27160 & 0.37827 -i 0.27160 \\
   & \mathrm{`` \infty "}& 0.37827 - i 0.27160 & \\
\hline
\end{array}
\]
\caption{Quasi-normal mode frequencies computed at different orders $k$ of the Taylor series for $V_\mathrm{PT}$. 
The $\omega^\mathrm{Exact}_{k}$ shows the exact value computed analytically, while $\omega^\mathrm{Numerical}_{k}$ shows the numerically computed one, at each Taylor order $k$ respectively. 
These results are for a step-size $\epsilon = 0.1$ and 11 point stencils. 
The true value using the exact analytic formula Eq.~\eqref{omega_PT} is labeled with $\mathrm{`` \infty "}$.}
\label{w_PT_table}
\end{table}

\subsection{Breit-Wigner Potential}\label{V_BW}

As next application we consider the potential
\begin{align}\label{BW_potential}
V_\text{BW}(x) = \frac{V_0}{1+(ax)^2},
\end{align}
which we call the Breit-Wigner (BW) potential. 
It has a similar shape as the PT potential around $x=0$, but different asymptotic behaviours. 
The transformations for $(V_0, a)$ are the same as for $(V_0, \alpha)$. 
To the best of our knowledge it has not been applied to black hole quasi-normal modes and we are not aware of the analytic form of the spectrum, but it might in principle exist in closed form. 
These aspects make the BW potential an attractive case for the numerical method. 

\begin{figure}
\centering
\includegraphics[width=1.0\linewidth]{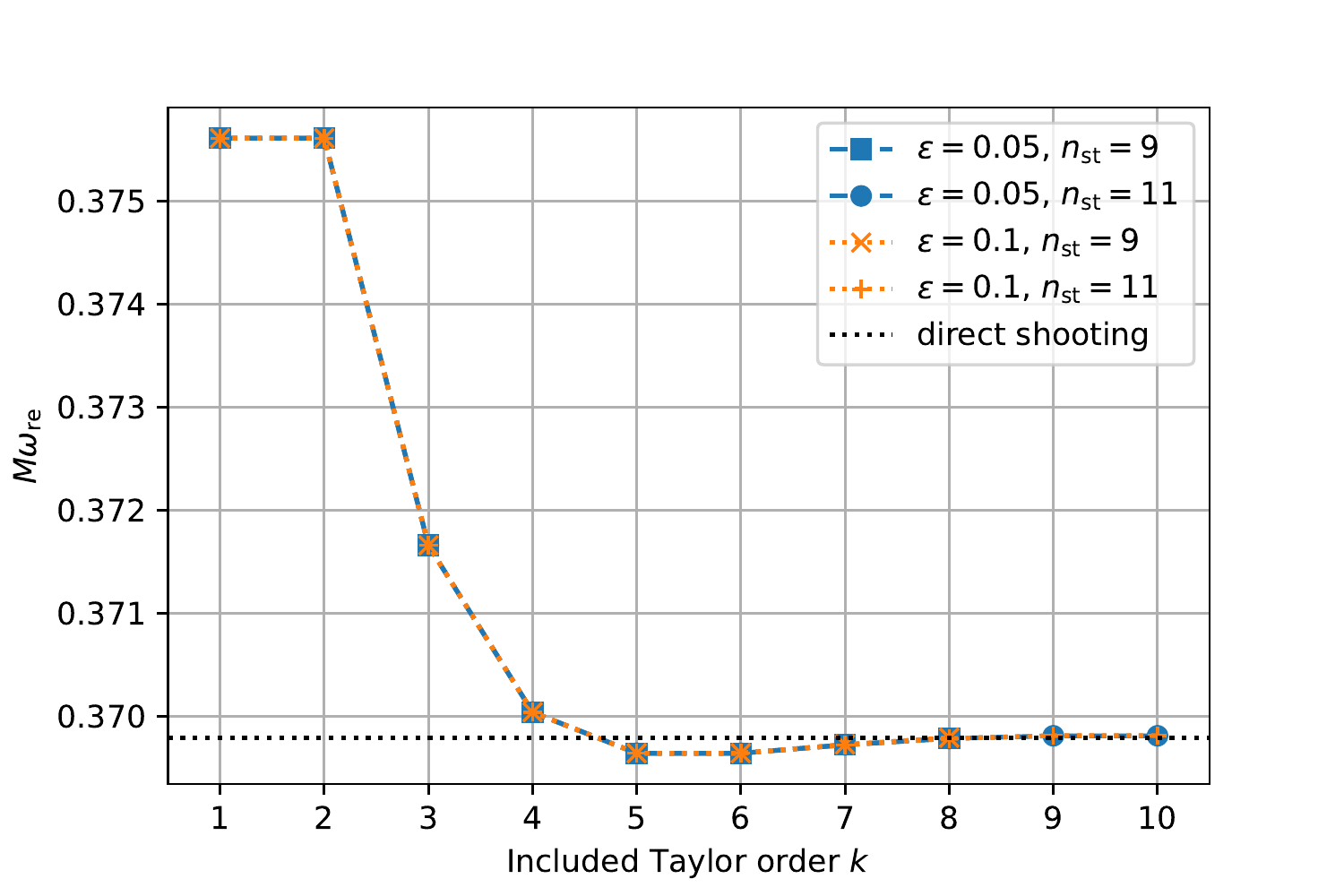} 
\includegraphics[width=1.0\linewidth]{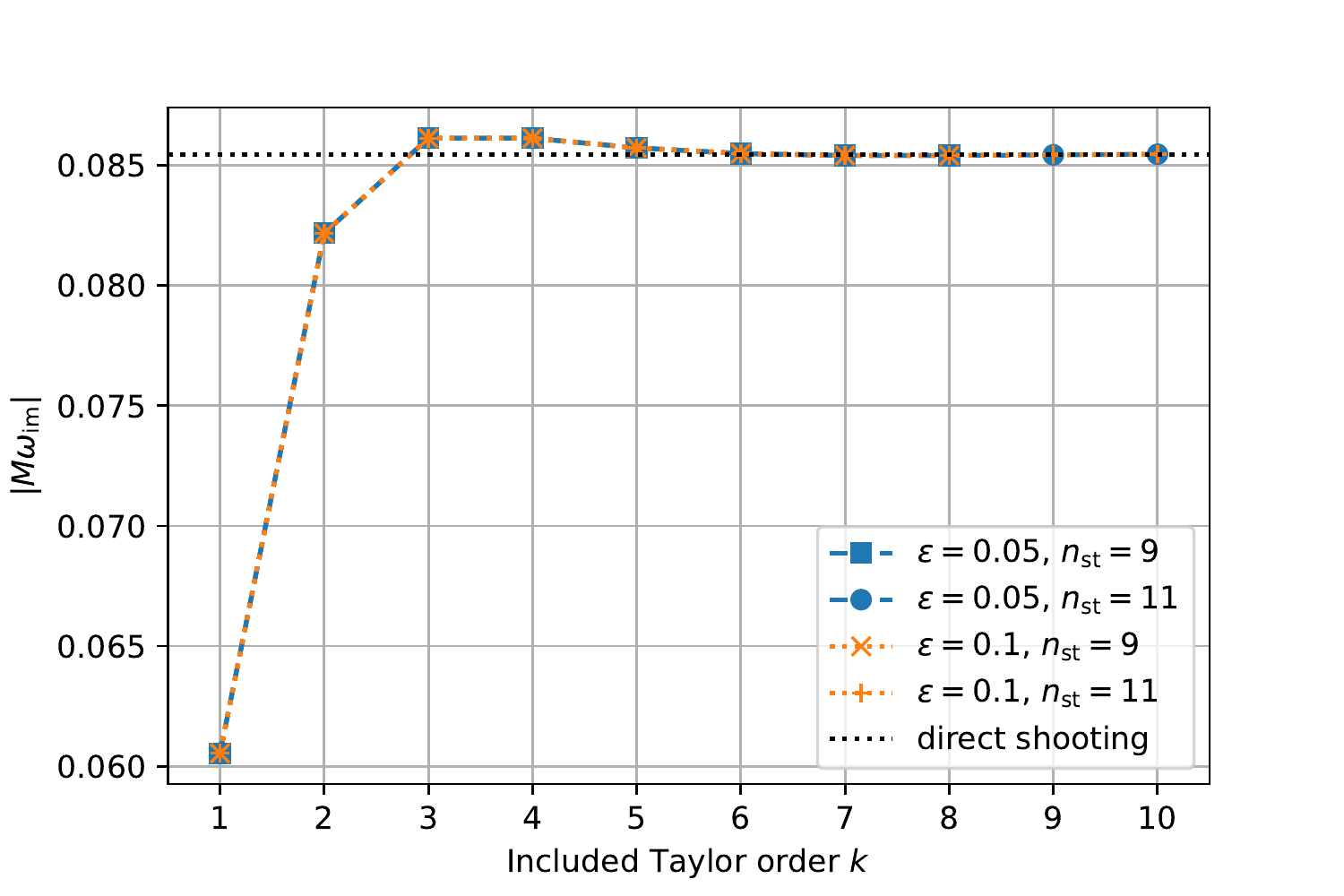} 
\caption{Here we show the real part (left panel) and imaginary part (right panel) for the $n=0$ quasi-normal mode computed for the BW potential for different step-sizes $\epsilon = (0.05, 0.1)$ and stencils of $n_\mathrm{st}=(9,11)$. 
The direct shooting values are indicated as black dotted line.}
\label{fig_BW_w_0}
\end{figure}

\begin{figure}
\centering
\includegraphics[width=1.0\linewidth]{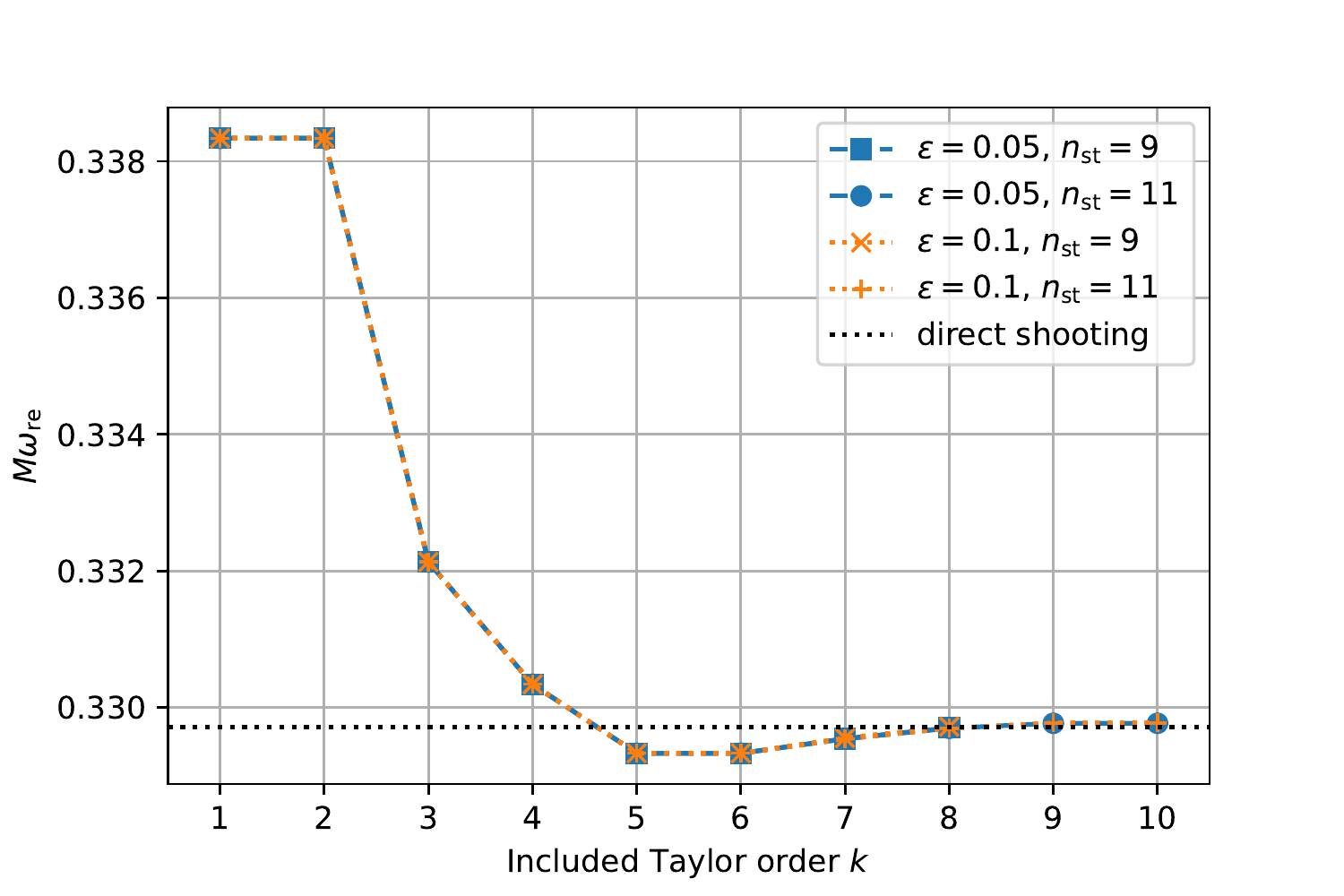}  
\includegraphics[width=1.0\linewidth]{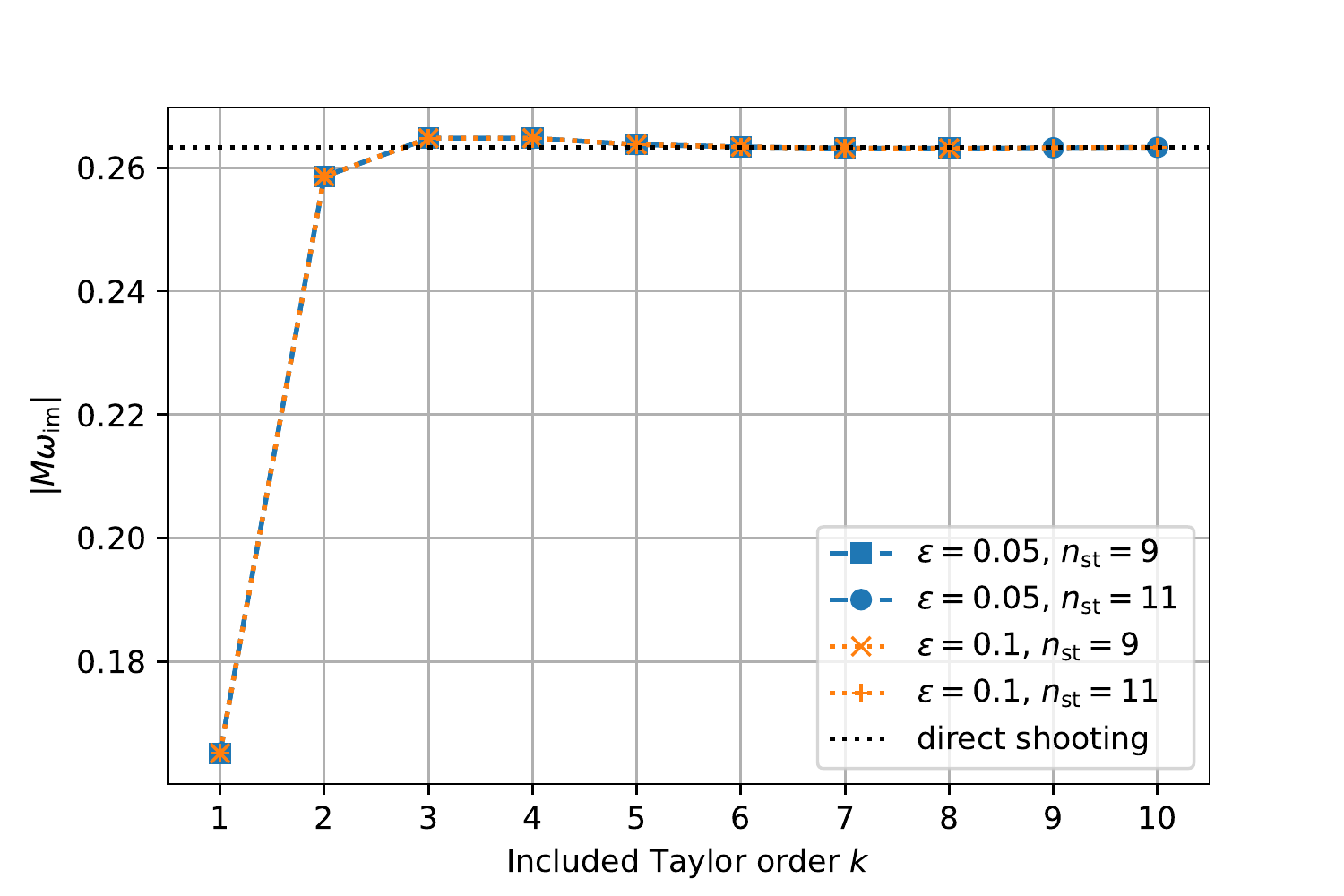} 
\caption{Here we show the real part (left panel) and imaginary part (right panel) for the $n=1$ quasi-normal mode computed for the BW potential for different step-sizes $\epsilon = (0.05, 0.1)$ and stencils of $n_\mathrm{st}=(9,11)$. 
The direct shooting values are indicated as black dotted line.}
\label{fig_BW_w_1}
\end{figure}

The values for different orders of the Taylor series are listed in Table \ref{w_BW_table} and compared to the other potentials in Fig.~\ref{fig_omega_all}. 
It can be seen that the convergence is not as fast as for the PT potential. 
We also show the quasi-normal modes computed with the direct shooting method for comparison, which demonstrates that the bound state method converges towards the correct values. 
Similar to the PT potential, we also show the convergence of the quasi-normal mode frequencies in Fig.~\ref{fig_BW_w_0} and Fig.~\ref{fig_BW_w_1} for different number of stencil points and step-sizes. 
Here we find excellent agreement for the real and imaginary part at all Taylor orders for the different numerical details. 

\begin{table}
\[
\begin{array}{ccc}
\hline 
 n & k & M \omega^\mathrm{Numerical}_{k}  \\
\hline
0 & 1  & 0.37561 -i 0.06054 \\
  & 2  & 0.37561 -i 0.08218 \\
  & 3  & 0.37166 -i 0.08613 \\
  & 4  & 0.37004 -i 0.08613 \\
  & 5  & 0.36964 -i 0.08573 \\
  & 6  & 0.36964 -i 0.08549 \\
  & 7  & 0.36972 -i 0.08541 \\
  & 8  & 0.36978 -i 0.08541 \\
  & 9  & 0.36981 -i 0.08544 \\
  & 10 & 0.36981 -i 0.08547 \\
 & \text{direct shooting}  & 0.36979 -i 0.08546 \\
\hline 
1 & 1  & 0.33834 -i 0.16514 \\
  & 2  & 0.33834 -i 0.25859 \\
  & 3  & 0.33213 -i 0.26479 \\
  & 4  & 0.33034 -i 0.26479 \\
  & 5  & 0.32933 -i 0.26378 \\
  & 6  & 0.32933 -i 0.26338 \\
  & 7  & 0.32954 -i 0.26317 \\
  & 8  & 0.32970 -i 0.26317 \\
  & 9  & 0.32978 -i 0.26325 \\
  & 10 & 0.32978 -i 0.26333 \\
 & \text{direct shooting} & 0.329(71) -i0.263(29) \\
\hline 
\end{array}
\]
\caption{Quasi-normal mode frequencies $\omega^\mathrm{Numerical}_{k}$ computed with the numerical method at different orders $k$ of the Taylor series for $V_\mathrm{BW}$. 
These results are for a step-size $\epsilon = 0.1$ and 11 point stencils. 
The numerical values obtained with the direct shooting method are shown for comparison and brackets indicate the accuracy limit. 
}
\label{w_BW_table}
\end{table}

\subsection{Mixed P\"oschl-Teller and Breit-Wigner Potential}\label{V_MIX}

As shown in Fig.~\ref{V_plots}, neither the popular PT potential nor the BW potential give a good approximation for both asymptotic behaviors for large $\pm x$. 
A more accurate matching can be obtained by combining the BW potential for $x \rightarrow \infty$ and the PT potential for $x \rightarrow -\infty$ as follows
\begin{align}\label{MIX_potential}
V_\text{MIX} (x) =
  \Big\{\begin{array}{lr}
        V_\text{PT}(x), \quad &\text{for } x < 0, \\
        V_\text{BW}(x), \quad &\text{for } x >= 0.
        \end{array}
\end{align}
Using the same $V_0$ in both potentials and setting $a = \alpha$ reduces the four dimensional Taylor series to one for only one non-trivial parameter $a$. 
The potentials are smoothly matched at $x=0$ due to the Taylor expansion $\cosh^2(\alpha x) = 1 + (\alpha x)^2 + \mathcal{O}((\alpha x)^4)$ around $\alpha x=0$. 

\begin{figure}
\centering
\includegraphics[width=1.0\linewidth]{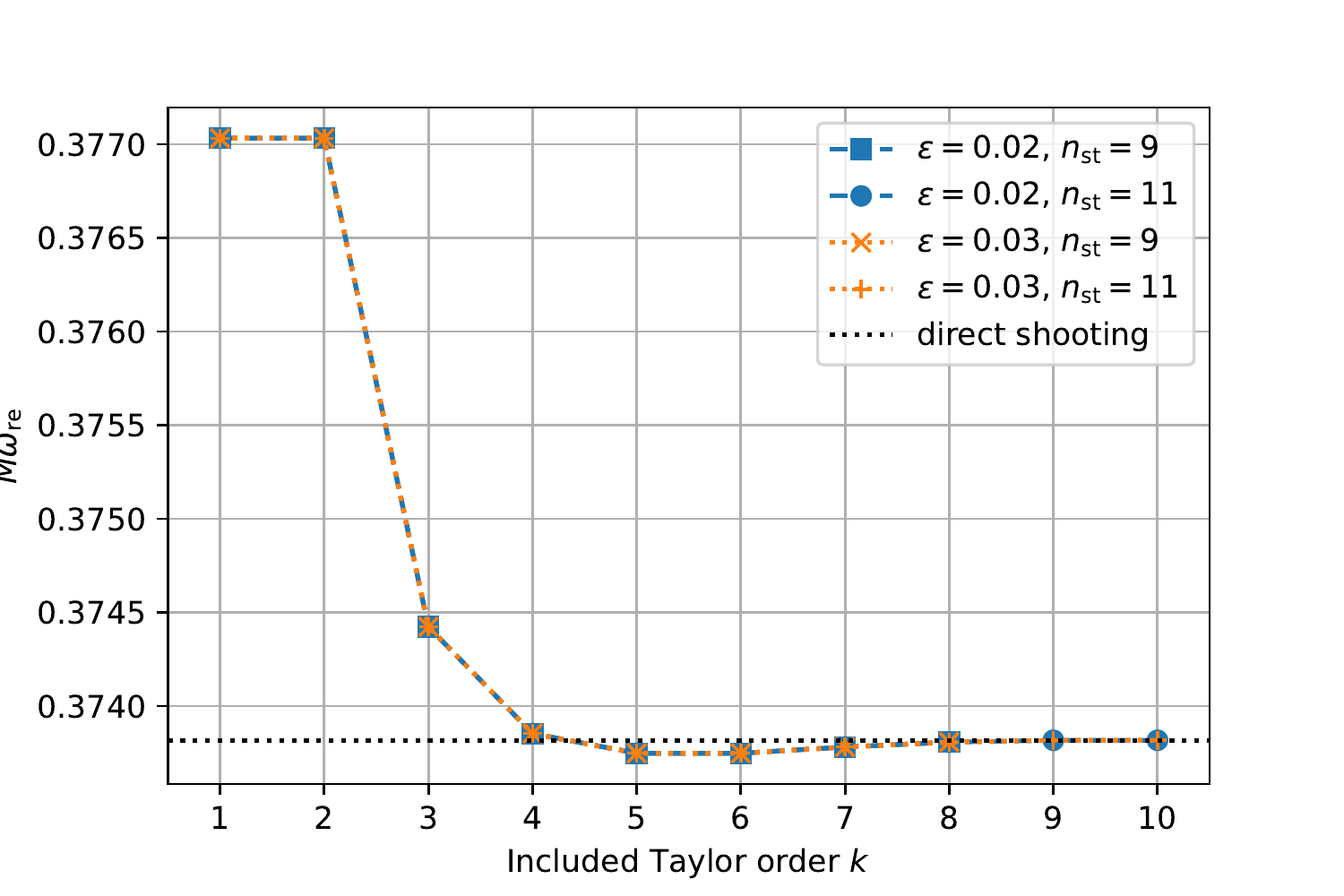}  
\includegraphics[width=1.0\linewidth]{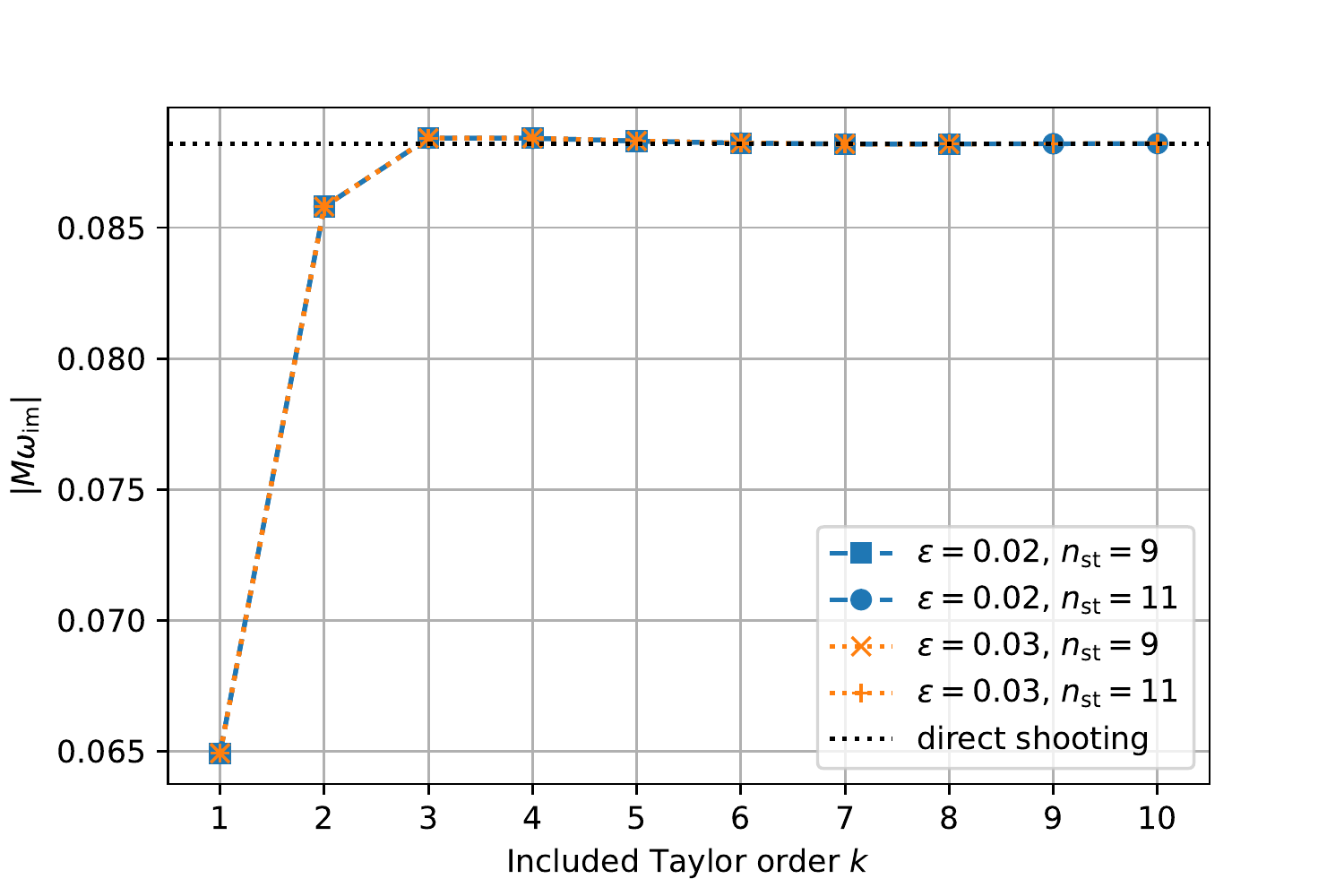} 
\caption{Here we show the real part (left panel) and imaginary part (right panel) for the $n=0$ quasi-normal mode computed for the mixed potential for different step-sizes $\epsilon = (0.05, 0.1)$ and stencils of $n_\mathrm{st}=(9,11)$. 
The direct shooting values are indicated as black dotted line.}
\label{fig_MIX_w_0}
\end{figure}

\begin{figure}
\centering
\includegraphics[width=1.0\linewidth]{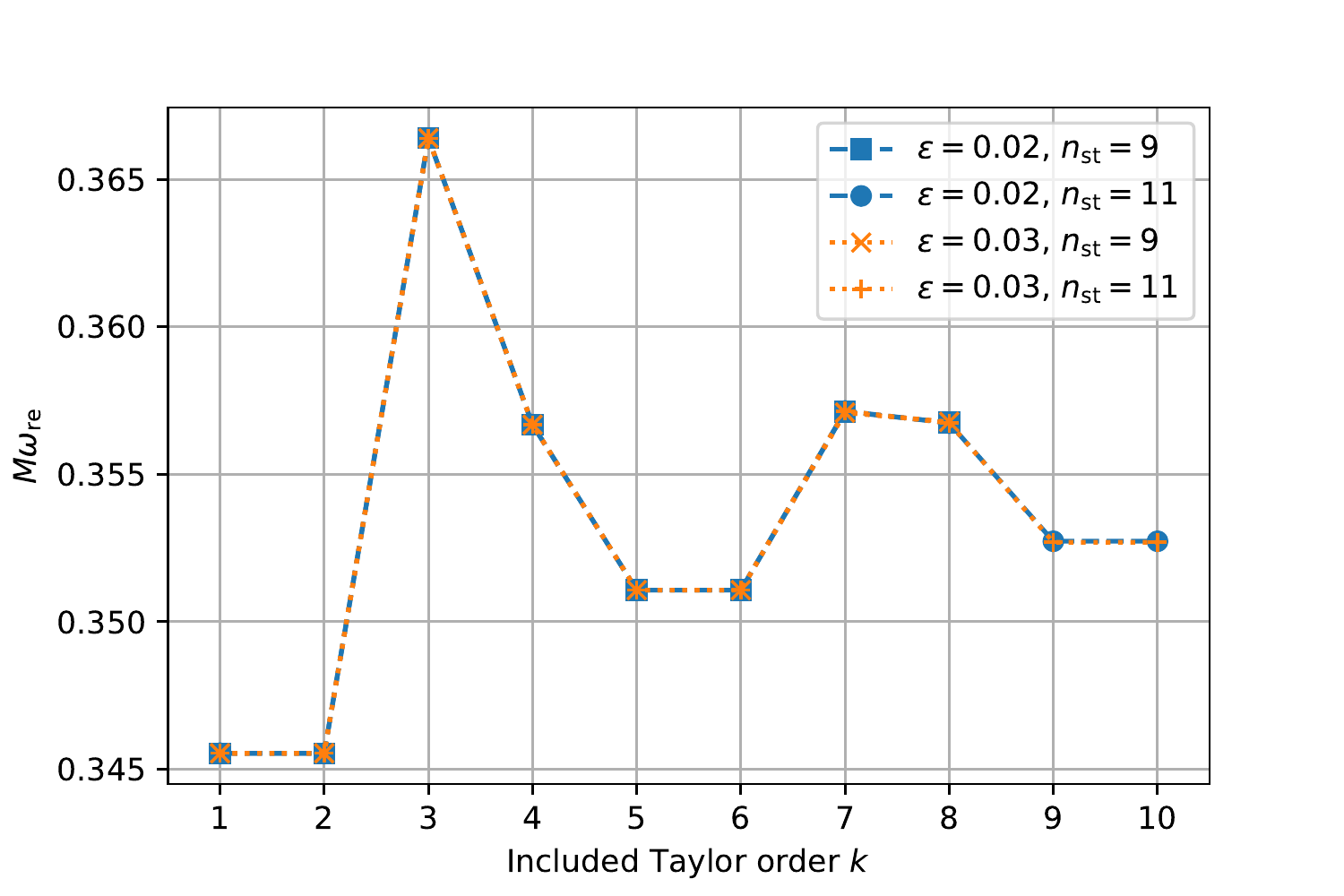} 
\includegraphics[width=1.0\linewidth]{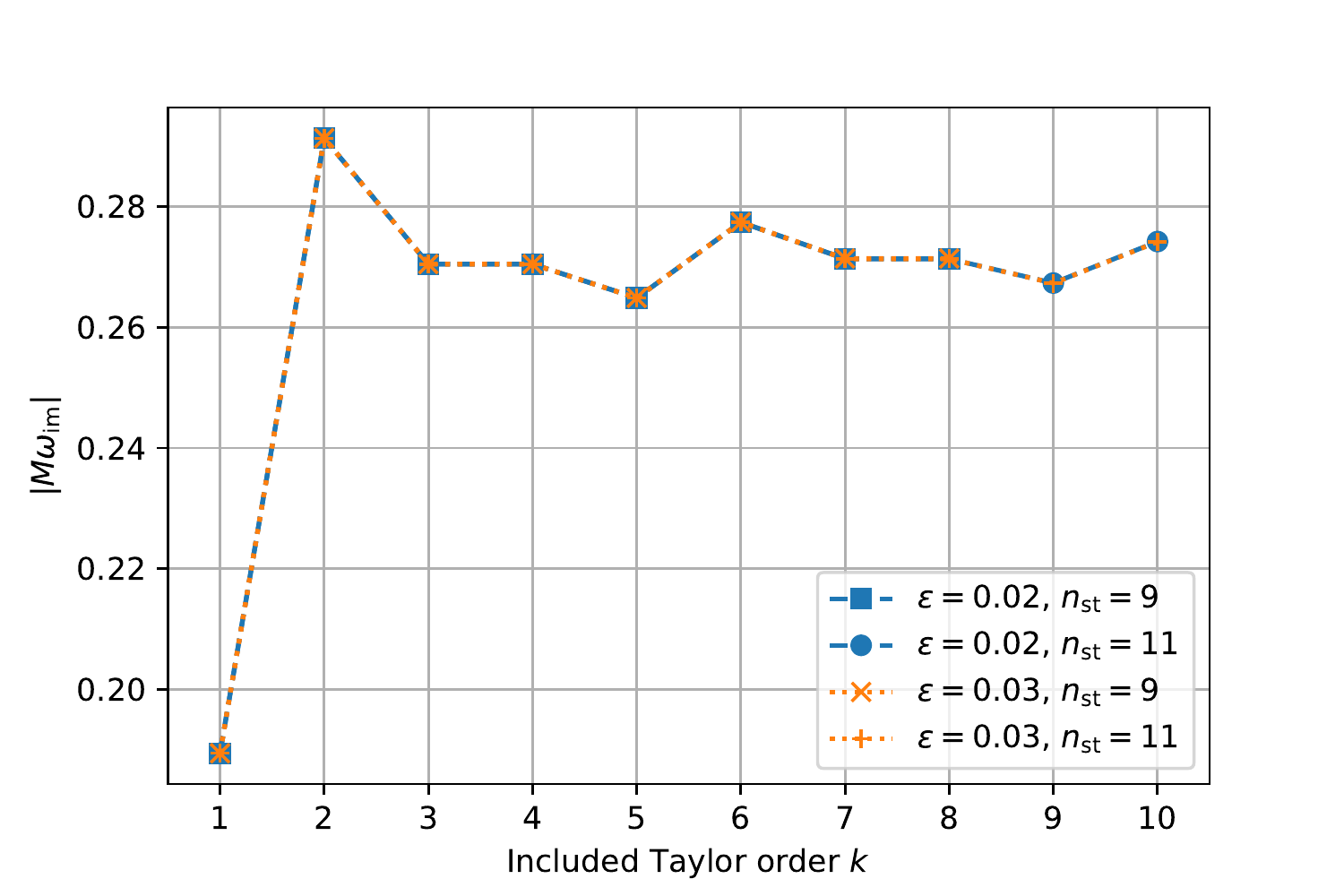}  
\caption{Here we show the real part (left panel) and imaginary part (right panel) for the $n=1$ quasi-normal mode computed for the mixed potential for different step-sizes $\epsilon = (0.05, 0.1)$ and stencils of $n_\mathrm{st}=(9,11)$. }
\label{fig_MIX_w_1}
\end{figure}

The results of the numerical method are presented in Table~\ref{w_MIX_table} and compared to the other potentials in Fig.~\ref{fig_omega_all}. 
We also show the $n=0$ quasi-normal mode computed with the direct shooting method for comparison, which demonstrates that the bound state method converges towards the correct values. 
Unfortunately it was not possible to robustly compute the $n=1$ quasi-normal mode via direct shooting with satisfying accuracy, so we do not report a value for comparison. 
Similar to the BW potential the convergence is slower compared to the one of the PT potential. 
Note that for $n=1$ the convergence is much slower, and in fact, more complicated. 
This type of ``oscillatory'' behaviour can be clearly seen in the bottom panels of Fig.~\ref{fig_MIX_w_0} and Fig.~\ref{fig_MIX_w_1}. 
This feature is robust with respect to changes in the stencils and step-sizes. 

\begin{table}
\[
\begin{array}{ccc}
\hline
 n & k & M \omega^\mathrm{Numerical}_{k}  \\
 \hline
0 & 1  & 0.37703 -i 0.06494 \\
  & 2  & 0.37703 -i 0.08580 \\
  & 3  & 0.37442 -i 0.08841 \\
  & 4  & 0.37385 -i 0.08841 \\
  & 5  & 0.37375 -i 0.08831 \\
  & 6  & 0.37375 -i 0.08822 \\
  & 7  & 0.37378 -i 0.08819 \\
  & 8  & 0.37381 -i 0.08819 \\
  & 9  & 0.37382 -i 0.08820 \\
  & 10 & 0.37382 -i 0.08821 \\
 & \text{direct shooting}  & 0.373(81) -i0.088(21) \\
\hline
1 & 1  & 0.34554 -i 0.18942 \\
  & 2  & 0.34554 -i 0.29132 \\
  & 3  & 0.36639 -i 0.27047 \\
  & 4  & 0.35668 -i 0.27047 \\
  & 5  & 0.35107 -i 0.26486 \\
  & 6  & 0.35107 -i 0.27741 \\
  & 7  & 0.35714 -i 0.27134 \\
  & 8  & 0.35674 -i 0.27134 \\
  & 9  & 0.35269 -i 0.26730 \\
  & 10 & 0.35269 -i 0.27416 \\
 & \text{direct shooting}  & - \\
\hline
\end{array}
\]
\caption{Quasi-normal mode frequencies $\omega^\mathrm{Numerical}_{k}$ computed with the numerical method at different orders $k$ of the Taylor series for $V_\mathrm{MIX}$. 
These results are for a step-size $\epsilon = 0.03$ and 11 point stencil. 
The numerical value for $n=0$ obtained with the direct shooting method is shown for comparison and brackets indicate the accuracy limit. 
}
\label{w_MIX_table}
\end{table}

\section{Discussion}\label{discussion}

\subsection{Comparison to exact Regge-Wheeler Quasi-Normal Modes}\label{V_ALL}

\begin{figure}
\includegraphics[width=1.0\linewidth]{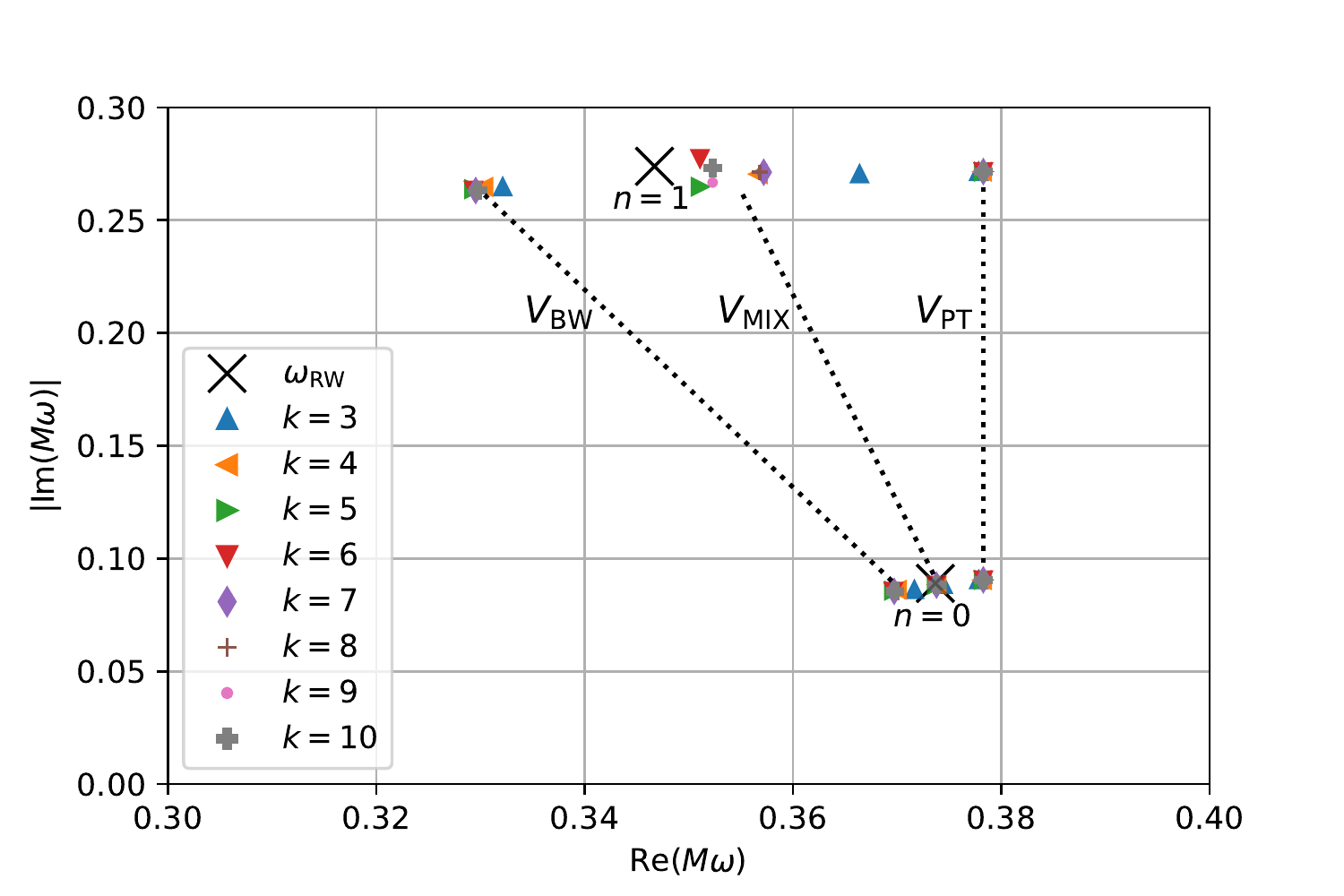}
\caption{Here we compare the $n=0$ and $n=1$ quasi-normal modes of the RW potential (black $+$) for $l=2$ with those of the PT potential (blue), BW potential (orange) and mixed potential (green) computed at different orders of the Taylor expansion (different marker types). 
Note that in some cases, especially for the PT potential, the markers overlap, demonstrating quick convergence. 
All values are also provided in Table~\ref{w_PT_table}, Table~\ref{w_BW_table} and Table~\ref{w_MIX_table}. 
The RW quasi-normal modes have been taken from supplementary material of Refs.~\cite{Berti:2005ys,Berti:2009kk}. 
}
\label{fig_omega_all}
\end{figure}

In Fig.~\ref{fig_omega_all} the exact quasi-normal modes of the RW potential are compared with those of the matched potentials studied in the previous sections. 
The figure shows the complex plane and the value of $\omega_k$ for different Taylor orders $k$. 
As can be clearly seen, the predicted values for the mixed potential are a much better approximation of the exact RW quasi-normal modes. 
Although the PT potential and the BW potential both give an adequate prediction for the $n=0$ mode, they fail for the real part of the $n=1$ mode. 
Note that the analytic result of the PT potential does not predict any change in the real part, independent of $n$, while the BW potential provides the correct trend. 
It is further remarkable, although somehow expected, that the quasi-normal modes of the mixed potential should line up between those of the PT potential and the BW potential. 
For $n=0$ the mixed potential approximation is excellent, while for $n=1$ the convergence of the Taylor series is not as fast as those of the other potentials. 

Can one also use the inverted potential approach to compute quasi-normal modes with $n\geq 1$? 
Interestingly the answer depends on what approximate potential is being used. 
In the case of the PT potential, there are only two bound states when matching to the $l=2$ RW potential because $\Omega_n$ changes sign between $n=1$ and $n=2$. 
The BW potential and mixed potential can admit more bound states, however those approach $0$ rather quickly and therefore their separation becomes very small. 
Because those states are less localized, the left and right starting points for the shooting method need to be increased significantly. 
This makes it numerically more difficult to compute the higher order derivatives with high accuracy. 

\subsection{Further Remarks on Accuracy}

The accuracy of the numerical bound state method depends on a couple of aspects. 
Obviously one cannot expect to get exactly the correct quasi-normal modes if approximate potentials are being used, which has been quantified in the original works \cite{Mashhoon:1982im,BLOME1984231,Ferrari:1984ozr,Ferrari:1984zz}. 
However, in the application to the RW potential, the mixed potential provides a significantly more accurate result than the PT potential. 
In the following we address the aspects particular to the method. 

First we want to comment on the numerical aspects. 
The accuracy of the method to compute the bound states at different points in the parameter space, which is then used to obtain the higher order derivatives in the Taylor series, can be identified as one technical bottleneck. 
Although bound states themselves can usually be computed with great accuracy with the shooting method, computing high derivatives from it correctly can be a non-trivial problem in practice, especially for overtones as mentioned in Sec.~\ref{V_ALL}. 
However, this can in principle be addressed by increasing the float precision in programming languages supporting it or in commercial software like \textsc{Mathematica} or \textsc{Maple}. 
Therefore, because standard tools exist and it is a widely studied topic, it might only be a problem if the quasi-normal modes need to be computed with arbitrary precision, for which other methods, e.g. the Leaver method \cite{Leaver:1985ax}, are better suited. 

To verify that the numerical results for the high derivatives are correct, we have checked several aspects of convergence, which we qualitatively summarize in the following. 
In particular we studied the initial starting points for the shooting method, different number of points for the finite differences (9, 11) and the corresponding step-size in the parameter used for the Taylor expansion. 
As expected we find that the overall accuracy in our tests increases for very high derivatives when more stencil points are included (for the same step-size). 
When using a large number of stencil points and qualitatively increasing the parameter step-size (which means one is less limited by the numerical accuracy of the bound states), one needs to be careful that one really computes the eigenvalue for the same overtone $n$ at all stencil points and does not jump to another overtone eigenvalue. 
Luckily these cases are very easy to exclude, because they predict very erratic results, and can be avoided by modifying the initial guess at each stencil point. 
Another test to verify the accuracy of the results is that we compared it to the analytic results, in case of the PT potential, and where possible to numerical results obtained via direct shooting for the BW and mixed potentials. 
We find very good agreement between the two methods and that the series seem to converge, within numerical uncertainties, to the expected values. 

The second aspect is on the analytic side. 
Even if many higher order derivatives can be computed accurately, it is in general not guaranteed that the Taylor series is valid at $\pi^{-1}(P)$, since this point might not even be close on the complex plane. 
In such a case, a clear warning sign would be that the quasi-normal modes do not converge as more terms are included. 
From a very qualitative point of view this could happen if the shape of the potential changes in a rather non-trivial way as function of $P$, which is not the case for any of the potentials studied in this work. 
The opposite case, which is that the series converges to some value, is of course not a rigorous proof that it converges to the right one. 
Comparing the Taylor series convergence of the quasi-normal modes of the different potentials used in this work, it can be seen that it is rapid for $n=0$ for all potentials. 
Looking at the first overtone $n=1$, the quasi-normal mode of the PT potential and the BW potential convergence rapidly, while the one of the mixed potential shows more variability, in particular for its real part.

\subsection{Insights from WKB Theory}\label{disc_WKB}

Although the better performance of the mixed potential could already be expected from Fig.~\ref{V_plots}, one can further understand it in terms of analytic properties. 
In particular insightful are results from WKB theory in the form of the classical Bohr-Sommerfeld quantization rule
\begin{align}\label{BS_rule}
\int_{x_0(E_n)}^{x_1(E_n)} \sqrt{E_n -  V(x)} \text{d} x = \pi \left(n+\frac{1}{2} \right).
\end{align}
Here $n\in \mathbb{N}_0$ labels the bound states and the turning points $x_0, x_1$ are defined as the root of the integrand (and thus dependent on $E_n$ themselves). 
It can be shown that potentials with the same separation of turning points, defined as
\begin{align}
L(E) \equiv x_1(E) - x_0(E),
\end{align}
have the same spectrum of bound states when computed with Eq.~\ref{BS_rule}, see. e.g. Refs.~\cite{Poschl:1933zz,lieb2015studies,MR985100,1980AmJPh..48..432L,Bonatsos:1992qq}. 
A similar result can also be found when studying the transmission through a two turning point potential barrier by inverting the Gamow formula \cite{lazenby1980classical,2006AmJPh..74..638G}, or even more general for certain three and four turning point potentials \cite{Volkel:2017kfj,Volkel:2018czg,Volkel:2018hwb}. 

\begin{figure}
\includegraphics[width=1.0\linewidth]{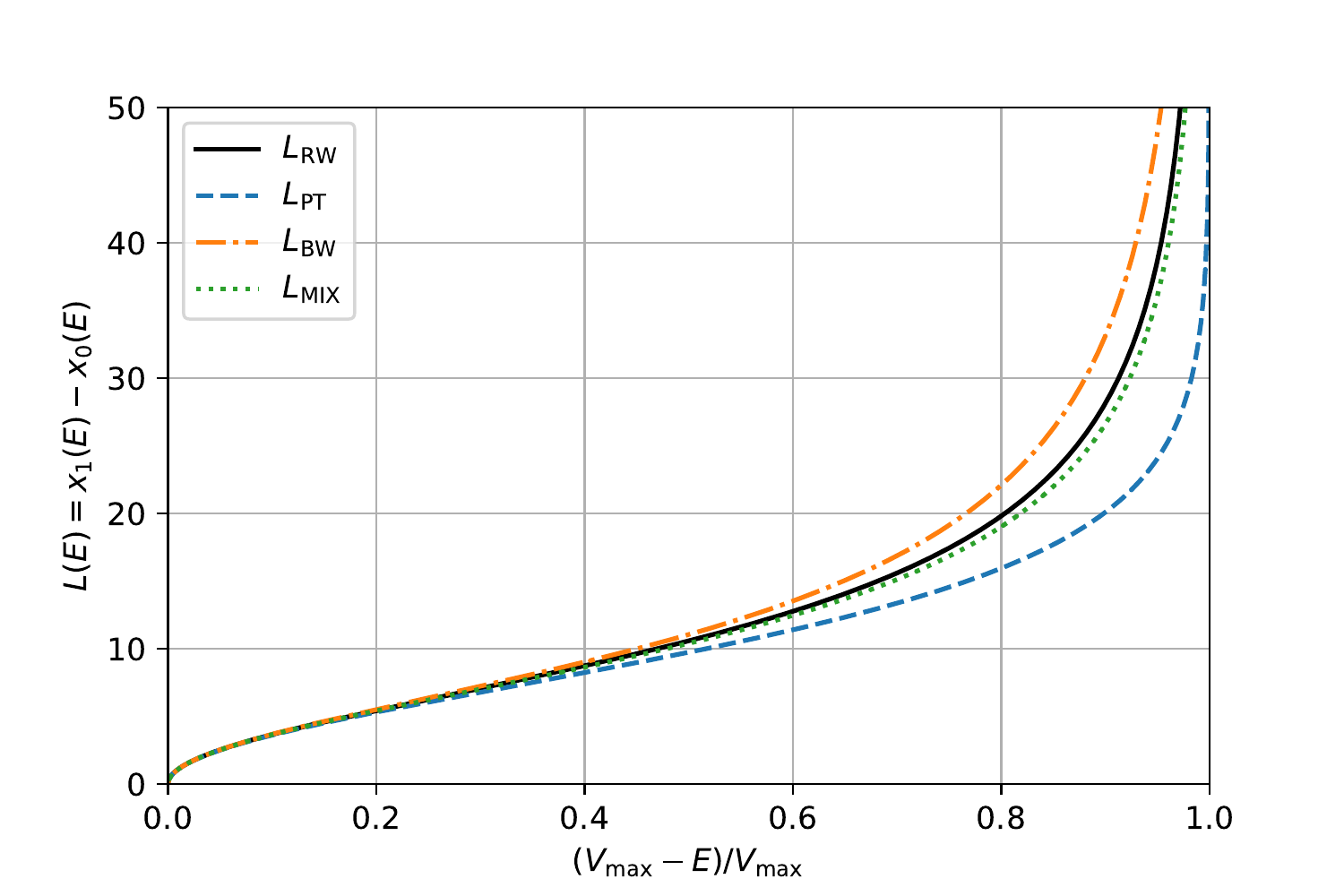}
\\
\includegraphics[width=1.0\linewidth]{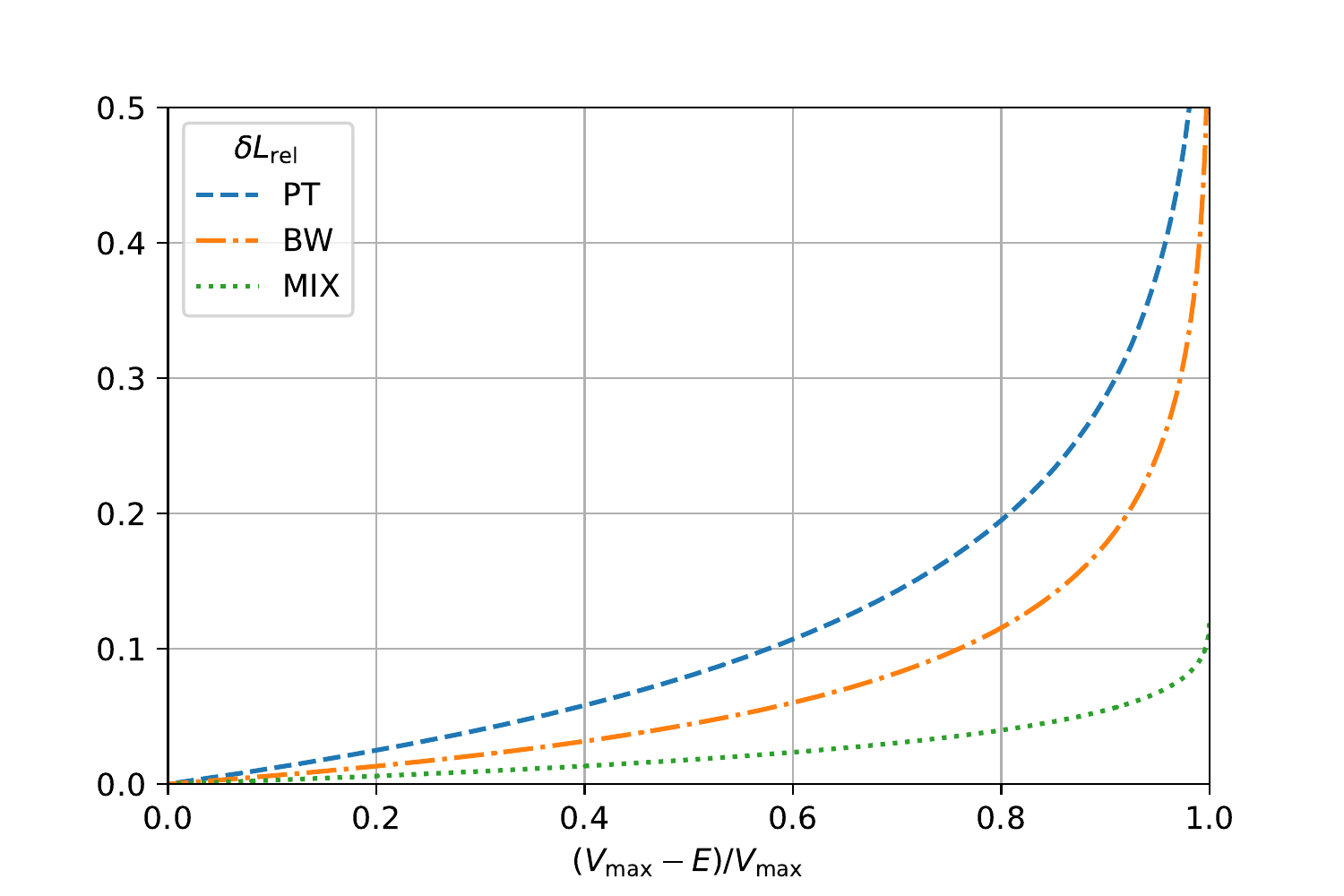}
\caption{Here we compare the widths of the RW potential for $M =1$ and $l=2$ (black) with those of the PT potential (blue), the BW potential (orange) and the mixed potential (green). As in Fig.~\ref{V_plots} the parameters of the approximate potentials have been chosen to agree with the RW at its peak. In the top panel we show each width, while the bottom panel shows the relative difference of a given width with respect to those of the RW potential.}
\label{L_plots}
\end{figure}

To further illustrate how this relates to the quasi-normal mode problem, we show $L$ for the different potentials in the top panel of Fig.~\ref{L_plots}, along with the relative error with respect to the RW potential in the bottom panel. 
As evident, all potentials are a good description close to the maximum/minimum of the potential, but deviate further away in a different way. 
Because quasi-normal modes are equivalent to bound states, which can be approximated by Eq.~\eqref{BS_rule}, one should expect that all three potentials perform well to describe the $n=0$ modes, but less good for $n=1$. 
Comparing the relative errors of the different potential widths in the bottom panel of Fig.~\ref{L_plots}, one would further expect for $n=1$ that the PT potential provides a bit worse approximation than the BW potential, but the mixed potential should be much better. 
This is indeed the case, as confirmed in Fig.~\ref{fig_omega_all}, or by comparing the different tables. 

\section{Conclusions}\label{conclusions}

The idea to obtain quasi-normal modes from the knowledge of bound states of the inverted potential is elegant and has been widely used since the mid 1980's \cite{Mashhoon:1982im,BLOME1984231,Ferrari:1984ozr,Ferrari:1984zz,Churilova:2021nnc}. 
However, the approach is limited to a few exactly solvable potentials, because it requires the spectrum of bound states in analytic form. 
The parameters of these solvable potentials are in most applications fit/matched to more complicated potentials for which the quasi-normal modes are requested. 
Although transformations that make potentials eligible to the method also exist for more complicated cases, the requirement of the analytic bound state spectrum is a crucial bottleneck of the original method. 
Another existing extension is based on perturbative calculations around the potential minimum, see e.g. Refs.~\cite{Hatsuda:2019eoj,Zaslavsky:1991ug,Matyjasek:2019eeu}. 
In this work we have introduced a novel approach to compute quasi-normal modes from bound states fully numerically and applied it to new approximate potentials. 

The key idea behind the numerical approach is to construct the Taylor expansion of the bound state spectrum for a given set of parameters $P$. 
This can be done with standard methods to compute bound states and we used the shooting method. 
If the Taylor series is computed with enough terms and converges when evaluated at $\pi^{-1}(P)$, the quasi-normal modes are directly given. 

We demonstrated the performance and identified possible bottlenecks and pitfalls by applying it to multiple potentials. 
The method can provide quasi-normal modes with high precision and can in principle be made even more precise in a straightforward way by including higher order terms of the underlying Taylor series. 

As natural test case of black hole quasi-normal modes, we have applied the method to a new approximate potential that captures the asymptotic properties of the RW potential better than the standard PT potential. 
Using the numerical method, we have shown that the quasi-normal modes of the new potentials provide a significantly more precise description, even for overtones where the PT potentials fails for the real part. 

Even if many methods to compute quasi-normal modes already exist, the inverted potential approach is also interesting from a numerical point of view. 
Using standard approaches, like the shooting method to directly search for the complex quasi-normal modes, is often unstable and requires a careful treatment of the asymptotic properties, as well as root finding on the complex plane. 
In contrast, the numerical computation of bound states with the same method is stable and the root finding typically only on the real axis. 

Future work could explore the numerical higher order Taylor series for potentials with multiple parameters. 
Another possible extension is to study eigenvalue dependent potentials, as they emerge for quasi-normal modes of rotating black holes, or even systems of coupled wave equations, which are rather common beyond general relativity. 

\acknowledgments

SV wants to thank Kostas D. Kokkotas, Christian Kr\"uger, Kostas Glampedakis, Enrico Barausse, Nicola Franchini and Mario Herrero-Valea for useful discussions, and in particular Andrew Coates, Nicola Franchini, Mario Herrero-Valea and Yasuyuki Hatsuda for constructive comments on the manuscript. 
SV acknowledges financial support provided under the European Union's H2020 ERC Consolidator Grant ``GRavity from Astrophysical to Microscopic Scales'' grant agreement no. GRAMS-815673. 
This work was supported by the EU Horizon 2020 Research and Innovation Programme under the Marie Sklodowska-Curie Grant Agreement No. 101007855.

\bibliography{literature}

\appendix \label{appendix}

\section{Computing Bound States with the Shooting Method}\label{appendix_bound}

In the following we summarize the main idea of the popular shooting method to compute the spectrum of bound states numerically, see e.g. Ref.~\cite{Press2007} for more details. 
In this approach one uses the asymptotic behavior of the wave equation at large negative and large positive values of $x$, called $x_{-}$ and $x_{+}$ in the following, and chooses some value for $E$ as initial guess. 
The physical suitable boundary conditions for $E < V(x)$ are those of exponentially decaying functions and their precise asymptotic form depends on the details of the potential. 
Assuming that the potential goes to zero for $x\rightarrow \pm \infty$ the physical solutions for bound states $E_n$ are imposed by the boundary conditions
\begin{align}\label{bc_0}
\Psi^\mathrm{bc}_{{\pm}}(x) &\sim \exp\left( \mp \sqrt{ - E_n} x \right).
\end{align}
The minus sign of $\Psi^\mathrm{bc}(x)$ labels the solution for $x\rightarrow -\infty$ and the plus sign vice versa. 
Since $E_n$ is not known one can find it from a root finding problem as follows. 
For some initial guess $E$ one can use Eq.~\eqref{bc_0} and its first derivative to start integrating a function $\Psi_{{\pm}}(x, E)$ as initial value problem from large values of $x_{-}$ and $x_{+}$ to some intermediate point $x_\text{M}$, where one matches the two solutions. 
The matching condition can be checked by evaluating the Wronskian, defined as
\begin{align}\label{wronskian}
W(\Psi_{-},\Psi_{+})(x, E)  =  \Psi_{-} \Psi_{+}^\prime - \Psi_{-}^\prime \Psi_{+}.
\end{align}
If the chosen value of $E$ is a solution to the imposed boundary conditions Eq.~\eqref{bc_0}, the Wronskian will vanish at $x_\text{M}$, but will be non-zero otherwise. 
Considering the Wronskian as function of $E$ one can find the spectrum of eigenvalues $E_n$ by computing the roots of the Wronskian. 
This one-dimensional root finding problem can be done numerically. 

\section{Computing Quasi-Normal Modes with the Shooting Method}\label{appendix_qnm}

Computing quasi-normal modes with the shooting method is significantly more involved than computing bound states, again we refer the interested reader to Refs.~\cite{Kokkotas:1999bd,Nollert:1999ji,Berti:2009kk,Konoplya:2011qq,Pani:2013pma} for extensive reviews. 
We remind the reader that the asymptotic form of the desired solution for $\psi^\mathrm{bc}(x)$ (now referring to the quasi-normal mode problem) corresponds to boundary conditions of purely outgoing waves for $x\rightarrow \infty$ and purely ingoing waves at $x\rightarrow -\infty$
\begin{align}\label{bc_1}
\psi^\mathrm{bc}_{\pm}(x) \sim \exp\left(\pm i \omega_n x \right).
\end{align}
Because the quasi-normal mode spectrum $\omega_n$ is in general complex valued, and corresponds to exponentially growing solutions towards $x \rightarrow \pm \infty$, one cannot set the starting point for the shooting at arbitrarily large values of $x$. 
This is because the numerical integration towards the intermediate matching point $x_\text{M}$ where the Wronskian Eq.~\eqref{wronskian} (now instead using $\psi(x, \omega)$) is computed is unstable. 
This is partially due to the fact that the other, unwanted solution of an inwards exponentially growing wave is ``excited'' from numerical errors coming from the integration itself. 
This forces one to consider starting points that are closer to the potential barrier, but here the asymptotic solutions of purely outgoing/ingoing waves are not valid yet and thus the boundary conditions Eq.~\eqref{bc_1} are less accurate. 

A common way to circumvent this problem is to consider high order corrections of the form 
\begin{align}\label{bc_2}
\psi^\mathrm{bc}_{\pm}(x) \sim \exp\left(\pm i \omega_n x\right) \sum_{j=0}^{j_\mathrm{max}}  \frac{c^{\pm}_j}{x^j}.
\end{align}
Here the coefficients $c^{\pm}_j$ depend on the details of the potential barrier, e.g. the two parameters of the BW potential. 
They are determined by inserting Eq.~\eqref{bc_2} into the Schr\"odinger equation Eq.\eqref{schrodinger} and expanding it in powers of $1/x$, for $x\rightarrow \pm \infty$. 
Comparing the resulting terms at each $1/x$ power then allows one to find the explicit form of the coefficients. 
The improved boundary conditions can now be used to choose closer integration points and thus increase the accuracy of the shooting method. 
To obtain the quasi-normal modes that were used in this work we have worked with $j_\mathrm{max}=25$, which yields lengthy and uninformative expressions for the set of $c^{\pm}_j$ that we do not report here. 

One might ask whether going up to $j_\mathrm{max}=25$ is really needed to obtain quasi-normal modes with reasonable accuracy. 
While for $n=0$ using a smaller number of terms is sufficient, the numerical challenges increase tremendously for overtones. 
This is because their imaginary part is growing significantly, which implies that it becomes much more challenging to integrate the correct solution for the reasons outlined previously. 
While the direct shooting results for $n=0$ quoted in this work are very robust with respect to different starting points for the integration, we do observe less robust results for $n=1$. 
Therefore we do not necessarily expect the direct shooting results for $n=1$ to be much more accurate than those obtained with the numerical bound state method at high Taylor orders. 
We have indicated the variability in the results with brackets around the relevant digits in Table~\ref{w_BW_table} and Table~\ref{w_MIX_table}. 
This indicates the expected error in the direct shooting results. 

The circumstance that the boundary conditions must be taken with so much care for the direct shooting method, but can be chosen very simplistically for the bound state case (the correction terms are not needed), clearly demonstrates one main disadvantage of the direct shooting method. 
Moreover, the chosen potentials in this work are very simple, but finding a higher order series expansion for the boundary conditions may be non-trivial for more complicated potentials and thus not always possible in practice. 

\end{document}